\documentclass[aps,pra,twocolumn,showpacs,showkeys,preprintnumbers,amsmath,amssymb]{revtex4}

\usepackage{graphicx}
\usepackage{dcolumn}
\usepackage{bm}

\begin{document}

\preprint{APS/123-QED}

\title{Green-function method in the theory of ultraslow
       electromagnetic waves\\ in an ideal gas with
       Bose-Einstein condensates}

\author{Yurii Slyusarenko}
 \email{slusarenko@kipt.kharkov.ua}
\author{Andrii Sotnikov}%
\affiliation{%
Akhiezer Institute for Theoretical Physics, NSC KIPT,\\ 1
Akademichna Street, Kharkiv 61108, Ukraine
}%

\date{\today}

\begin{abstract}
    We propose a microscopic approach describing the
    interaction of an ideal gas of hydrogenlike atoms
    with a weak electromagnetic field. This approach is based on the
    Green-function formalism and an approximate formulation of the
    method of
    second quantization for quantum many-particle systems
    in the presence of bound states of particles.
    The dependencies of the propagation
    velocity and damping rate of electromagnetic pulses
    on the microscopic characteristics of the system
    are studied for a gas of hydrogenlike atoms.
    For a Bose-Einstein condensate of alkali-metal atoms
    we find the conditions when the electromagnetic waves
    of both the optical and microwave regions are slowed.
    In the framework of the proposed approach, the influence
    of an external homogeneous and static magnetic field
    on the slowing phenomenon is studied.
\end{abstract}

\pacs{03.75.Hh, 05.30.-d, 42.25.Bs}%
\keywords{atom-photon collisions, Bose-Einstein condensation, bound
states, Green's function methods, quantization (quantum theory)}

\maketitle

\section{Introduction}

The possibility of observation of ultraslow light pulses in vapors
of alkali-metal atoms at extremely low temperatures was shown in
1999~\cite{Hau1999N}. To the present moment, there are quite a lot
of theoretical works devoted to the description of this phenomenon
(probably the earliest of which is represented in
Ref.~\cite{Hau2004PRA}). Naturally, for the most complete
description of electromagnetic waves slowing in a Bose-Einstein
condensate (BEC), the development of an appropriate microscopic
approach is needed.

But, while constructing this approach, some problems appear. First
of all, one needs to use the method of second quantization for a
microscopic description of the response of the BEC to the
perturbation by the external electromagnetic field. The
second-quantization method uses the creation and annihilation
operators of atoms. This fact requires consideration of atoms as
elementary objects, which do not have internal structure. However,
atoms are compound objects with a rather complicated internal
structure. Moreover, in the processes of interaction of neutral
atoms with an electromagnetic field, this structure plays a crucial
role. In particular, it leads also to a sufficient reduction of the
group velocity of propagation of electromagnetic waves in a BEC.

Thus, a problem appears concerned with the introduction of the
creation and annihilation operators, which must preserve the
information about the internal structure and, first of all, about
the energy state of the atoms. In other words, we discuss the
formulation of the second-quantization method for a many-particle
system in the presence of bound states of particles (atoms). This
problem is solved in Ref.~\cite{Pel2005JMP}. In that work, an
approximate formulation of the second quantization method for such a
system was constructed in the case when the binding energy of
compound particles is much larger than the kinetic energy. The
possibility of development of such a theory is demonstrated by
considering a system that consists of fermions of two different
kinds and bound states (atoms) that can be formed only by two
fermions of different kinds (ions and electrons). The choice of this
model in Ref.~\cite{Pel2005JMP} was dictated not by methodological
difficulties, but to get more visual results. In that work, the
Hamiltonians of interaction for particles of different kinds
(including bound states) were constructed. The Hamiltonian of
interaction of all particles in the system (including neutral
compound particles) with the electromagnetic field was found.
Because of this, the authors constructed a nonrelativistic quantum
electrodynamics of systems which may consist of bound states of
particles. Within this theory, the Maxwell-Lorentz equations were
obtained. These allow study of the response of a system consisting
of charged \textquotedblleft elementary'' particles and bound states
to perturbation by an external electromagnetic field. For solving
this problem it is natural to use the Green-function formalism (see,
e.g., Refs.~\cite{Akhiezer1981,Sly2006CMP}).

In Ref.~\cite{Sly2006CMP} the Green-function method was generalized
to the case of description of this response of the systems mentioned
above. In terms of Green functions, the expressions for such
macroscopic characteristics as conductivity, permittivity, and
magnetic permeability of a hydrogenlike plasma were found.

In the present paper, the formulas that were found in
Refs.~\cite{Pel2005JMP,Sly2006CMP} are used to study the phenomenon
of electromagnetic pulses slowing in a BEC. According to the main
results from Refs.~\cite{Sly2006CMP,Sly2007LTP}, let us briefly
summarize some basic principles of the proposed approach.

\section{Hamiltonian of hydrogenlike atomic gas in an external
electromagnetic field}

Usually, BEC is studied in gases of alkali-metal atoms. It is known
that the internal structure of alkali metals is similar to the
internal structure of a hydrogen atom (hydrogenlike atoms). Thus,
such atoms may be considered as bound states of two elementary
particles (the valence electron and the atomic core). In this case,
the results of Refs.~\cite{Pel2005JMP,Sly2006CMP} become suitable
for a description of the interaction of a weak electromagnetic field
with a BEC. Let us note that in Ref.~\cite{Sly2006CMP} the Green
functions of hydrogenlike plasma were found. In other words, the
contribution of free fermions and bound states (atoms) to the Green
functions was considered. But the recent theoretical
investigations~\cite{Sly2007LTP} of a low-temperature hydrogenlike
plasma in the equilibrium state show that the density of free
fermions in such a system is exponentially small in comparison with
the atomic density. Therefore, at low temperatures one may neglect
the free-fermion contribution in the expressions for the Green
functions. Taking into account this fact, the system Hamiltonian may
be written as follows:
\begin{equation}\label{eq:2.1}                                    
    \hat{\mathcal H}(t)
    =\hat{\mathcal H}_0
    +\hat{\mathcal H}_{\text{int}}
    +\hat{V}(t),\quad
    \hat{\mathcal H}_0
    =\hat{\mathcal H}_{\text{ph}}
    +\hat{\mathcal H}_{\text{p}},
\end{equation}
where $\hat{\mathcal H}_{\text{ph}}$ is the Hamiltonian for free
photons and $\hat{\mathcal H}_{\text{int}}$ is the Hamiltonian of
interaction between atoms. Note that we neglect the
Hamiltonian~$\hat{\mathcal H}_{\text{ph}}$ below. We also consider
the system of hydrogenlike atoms as an ideal gas. Thus, we do not
take into account the presence of the interaction
Hamiltonian~$\hat{\mathcal H}_{\text{int}}$ in the next
calculations. Its microscopic expression is found in
Ref.~\cite{Pel2005JMP} in the framework of the developed formulation
of the second-quantization method. This Hamiltonian is constructed
on the basis of the Coulomb interaction between particles that can
form a bound state (hydrogenlike atom). The explicit form for it in
a low-energy approximation is given in the Appendix. We use it to
find the limits of applicability for the model of the ideal gas.

The operator~$\hat{\mathcal H}_{\text{p}}$ in Eq.~(\ref{eq:2.1}) is
the Hamiltonian for free particles (atoms),
\begin{equation}\label{eq:2.2}                                    
    \hat{\mathcal H}_{\text{p}}
    =\sum_{\alpha}\int d\textbf{x}
    \left(
    \dfrac{1}{2m}
    \dfrac{\partial{\hat\eta}_{\alpha}^{\dag}(\textbf{x})}
    {\partial\textbf{x}}
    \dfrac{\partial{\hat\eta}_{\alpha}
    (\textbf{x})}{\partial\textbf{x}}
    +\varepsilon_{\alpha}{\hat\eta}_{\alpha}^{\dag}
    (\textbf{x}){\hat\eta}_{\alpha}(\textbf{x})
    \right),
\end{equation}
where ${\hat\eta}_{\alpha}^{\dag}(\textbf{x})$ and
${\hat\eta}_{\alpha} (\textbf{x})$ are the creation and annihilation
operators of hydrogenlike (alkali-metal) atoms with the set of
quantum numbers~$\alpha$ at the point~$\textbf{x}$, respectively;
$\varepsilon_{\alpha}$ is the atomic energy corresponding to this
state; and $m$ is the atomic mass.

The operator~$\hat{V}(t)$ in Eq.~(\ref{eq:2.1}) represents the
Hamiltonian that describes the interaction of atoms with an
electromagnetic field,
\begin{equation}\label{eq:2.3}                                    
    \hat{V}(t)
    =-\dfrac{1}{c}\int d\textbf{x}~\textbf{A}(\textbf{x},t)
    \hat{\textbf{j}}(\textbf{x})
    +\int d\textbf{x}~\varphi(\textbf{x},t)
    \hat{\sigma}(\textbf{x}),
\end{equation}
where $\textbf{A}(\textbf{x},t)$ and $\varphi(\textbf{x},t)$ are the
vector and scalar potentials of the external electromagnetic field,
and $\hat{\textbf{j}}(\textbf{x})$ and $\hat{\sigma}(\textbf{x})$
are the operators of the current and charge densities for neutral
atoms, respectively (see Ref.~\cite{Pel2005JMP} for details),
\begin{eqnarray}                                                  
    \hat{\sigma}(\textbf{x})=\dfrac{1}{\mathcal{V}}
    \sum\limits_{\textbf{p},\textbf{p}'}
    \sum\limits_{\alpha,\beta}
    e^{i\textbf{x}(\textbf{p}'-\textbf{p})}
    \sigma_{\alpha\beta}(\textbf{p}-\textbf{p}')
    \hat{\eta}_{\alpha}^{\dag}(\textbf{p})
    \hat{\eta}_{\beta}(\textbf{p}')\label{eq:2.4},
    \\
    \hat{\textbf{j}}(\textbf{x})=\dfrac{1}{\mathcal{V}}
    \sum\limits_{\textbf{p},\textbf{p}'}
    \sum\limits_{\alpha,\beta}
    e^{i\textbf{x}(\textbf{p}'-\textbf{p})}
    \biggl(\textbf{I}_{\alpha\beta}(\textbf{p}-\textbf{p}')\nonumber
    \biggr.\qquad\qquad\quad
    \\
    \qquad\biggl.
    +\dfrac{(\textbf{p}+\textbf{p}')}{2m}
    \sigma_{\alpha\beta}(\textbf{p}-\textbf{p}')
    \biggr)
    \hat{\eta}_{\alpha}^{\dag}(\textbf{p})
    \hat{\eta}_{\beta}(\textbf{p}')\label{eq:2.5},
\end{eqnarray}
where $\mathcal{V}$ is the volume of the system. The matrix elements
of the charge and current densities in Eqs.~(\ref{eq:2.4}) and
(\ref{eq:2.5}) can be expressed in terms of the atomic wave
functions~$\varphi_{\alpha}(\textbf{x})$:
\begin{eqnarray}                                                  
    &\sigma_{\alpha\beta}(\textbf{k})&
    =e\int d\textbf{y}~\varphi_{\alpha}^{*}(\textbf{y})
    \varphi_{\beta}(\textbf{y})\nonumber
    \\
    &&\times\left[\exp{\left(i\dfrac{m_{\text{p}}}{m}
    \textbf{k}\textbf{y}\right)}
    -\exp{\left(-i\dfrac{m_{\text{e}}}{m}
    \textbf{k}\textbf{y}\right)}\right]\label{eq:2.6},
    \\
    &\textbf{I}_{\alpha\beta}(\textbf{k})&
    =\dfrac{ie}{2m_{\text{e}}}\int d\textbf{y}
    \left(\dfrac{\partial\varphi_{\alpha}^{*}(\textbf{y})}
    {\partial\textbf{y}}\varphi_{\beta}(\textbf{y})
    -\varphi_{\alpha}^{*}(\textbf{y})
    \dfrac{\partial\varphi_{\beta}(\textbf{y})}
    {\partial\textbf{y}}
    \right)\nonumber
    \\
    &&\times
    \left[\exp{\left(-i\dfrac{m_{\text{p}}}{m}
    \textbf{k}\textbf{y}\right)}
    +\dfrac{m_{\text{e}}}{m_{\text{p}}}
    \exp{\left(i\dfrac{m_{\text{e}}}{m}
    \textbf{k}\textbf{y}\right)}\right],\label{eq:2.7}
\end{eqnarray}
where $e$ is the electron charge absolute value, and $m_\text{p}$
and $m_\text{e}$ are the masses of the atomic core and electron,
respectively ($m=m_\text{p}+m_\text{e}$).

\section{Response of a gas of hydrogenlike atoms to perturbation
by an external electromagnetic field: Green functions}

Here we find the expressions that describe the influence of a weak
external electromagnetic field on the system under consideration. To
this end, following Ref.~\cite{Akhiezer1981}, let us recall some
main principles of the Green-function method.

Consider a system that in some moment of time $t$ is characterized
by a statistical operator~$\hat{\rho}(t)$. Then the system
Hamiltonian may be written as
\begin{equation*}
\hat{\mathcal H}(t)
    =\hat{\mathcal H}_0
    +\hat{V}(t),\quad \hat{V}(t)=\int
    d\textbf{x}~F_{i}(\textbf{x},t){\hat\xi}_{i}(\textbf{x}),
\end{equation*}
where $\hat{\mathcal H}_0$ is the Hamiltonian of an ideal atomic gas
and $\hat{V}(t)$ is the Hamiltonian of interaction of atoms with the
external electromagnetic field, $F_{i}(\textbf{x},t)$ are the
quantities, which define the external field, and
${\hat\xi}_{i}(\textbf{x})$ are the quasilocal operators, which
refer only to the studied system (they are not associated with the
external field; see Ref.~\cite{Akhiezer1981}). Note also that the
summation convention over the repeated index~$i$ is meant.

Let us consider below that the influence of the external field on
the system is weak. Due to this fact, we can develop perturbation
theory over~$\hat{V}(t)$. Thus, for the mean value of the arbitrary
quasilocal operator~$\hat{a}(\textbf{x})$ in the linear order
in~$\hat{V}(t)$, one gets
\begin{eqnarray}                                                  
    &&\text{Sp}\hat{\rho}(t){\hat a}(\textbf{x})
    =\text{Sp}w{\hat a}(0)+a^{F}(\textbf{x},t),\label{eq:3.1}
    \\
    &&a^{F}(\textbf{x},t)
    =\int_{-\infty}^{\infty}
    dt'\int d\textbf{x}'G_{a\xi_{i}}
    ({\textbf{x}}-\textbf{x}',t-t')
    F_{i}(\textbf{x}',t'),\nonumber
\end{eqnarray}
where $w$ is the equilibrium statistical operator (the Gibbs
operator) for the system under consideration,
\begin{equation}\label{eq:3.2}                                    
    w=\exp\left[\Omega-\beta
    \left(\hat{\mathcal H}_0
    -\sum\limits_{\alpha}\mu_{\alpha}
    \hat{N}_{\alpha}\right)\right].
\end{equation}
Here $\beta={1/T}$ is the reciprocal temperature in energy units,
$\mu_{\alpha}$ is the chemical potential of atoms in the
state~$\alpha$, and $\hat{N}_{\alpha} =\int
d\textbf{x}~\hat{\eta}^{\dag}_{\alpha}(\textbf{x})
\hat{\eta}_{\alpha}(\textbf{x})$ is the operator of the total number
of atoms in the state~$\alpha$. The temperature and the chemical
potential are found from the following equations:
\begin{equation*}                                                 
    \text{Sp}w{\hat{\mathcal H}_0}
    ={\mathcal H}_0,
    \quad\text{Sp}w\hat{N}_{\alpha}={N}_{\alpha}.
\end{equation*}
The dependence of the thermodynamic
potential~$\Omega=\Omega(T,\mu_{\alpha})$ is defined by
$\text{Sp}w=1$.

The function $G_{a\xi_{i}} ({\textbf{x}}-\textbf{x}',t-t')$
in~(\ref{eq:3.1}),
\begin{equation}\label{eq:3.3}                                    
    G_{a\xi_{i}}({\textbf{x}}-\textbf{x}',t-t')
    =-i\theta(t-t')\text{Sp}w[{{\hat
    a}}(\textbf{x},t),
    {\hat{\xi_{i}}}(\textbf{x}',t')],
\end{equation}
is the two-time retarded Green function of the
operators~$\hat{a}(\textbf{x},t)$, $\hat{\xi_{i}}(\textbf{x}',t')$,
which are taken in the Heisenberg representation,
\begin{equation*}                                                 
    {\hat a}(\textbf{x},t)=e^{i\hat{\mathcal H}_0t} {\hat
    a}(\textbf{x})e^{-i\hat{\mathcal H}_0t}.
\end{equation*}
The quantity~$\theta(t)$ in Eq.~(\ref{eq:3.3}) is the Heaviside function.

Taking the Fourier transforms of the quantities $a^{F}$ and $F_{i}$,
\begin{eqnarray*}                                                  
    a^{F}(\textbf{x},t)
    =\dfrac{1}{(2\pi)^{4}}
    \int d{\textbf{k}}~d\omega~
    e^{-i(t\omega-{\textbf{k}}
    {\textbf{x}})}a^{F}(\textbf{k},\omega),
    \\
    F_{i}(\textbf{x},t)=\dfrac{1}{(2\pi)^{4}}
    \int d{\textbf{k}}~d\omega~e^{-i(t\omega
    -{\textbf{k}}{\textbf{x}})}
    F_{i}(\textbf{k},\omega),
\end{eqnarray*}
one gets
\begin{equation}\label{eq:3.4}                                    
    a^{F}(\textbf{k},\omega)
    =G_{a\xi_{i}}({\textbf{k}},\omega)
    F_{i}(\textbf{k},\omega)
\end{equation}
with
\begin{equation}\label{eq:3.5}                                    
    G_{a\xi_{i}}({\textbf{k}},\omega)
    =\int_{-\infty}^{\infty}dt\int
    d\textbf{x}~e^{i(t\omega-{\textbf{k}}
    {\textbf{x}})}G_{a\xi_{i}}({\textbf{x}},t).
\end{equation}

Let us use the derived expressions for studying the response of a
gas of alkali-metal atoms to the external electromagnetic field. To
this end, we must replace the quantities~$F_{i}(\textbf{x},t)$ in
Eq.~(\ref{eq:3.1}) by the vector~$\textbf{A}(\textbf{x},t)$ and
scalar~$\varphi(\textbf{x},t)$ potentials of the external
electromagnetic field. The operators~${\hat a}(\textbf{x},t)$ and
${\hat{\xi_{i}}}(\textbf{x}',t')$ must be replaced in this case by
the operators of the current density~$\hat{\textbf{j}}(\textbf{x})$
and charge density~$\hat{\sigma}(\textbf{x})$ (see
Eqs.~(\ref{eq:2.3})--(\ref{eq:2.5})). By the
Hamiltonian~${\hat{\mathcal H}_0}$ in Eq.~(\ref{eq:3.2}) one must
take the Hamiltonian~$\hat{\mathcal H}_{\text{p}}$ (see
Eqs~(\ref{eq:2.1}) and (\ref{eq:2.2})). In next calculations we use
also the Maxwell-Lorentz equations for the average fields in the
system (see in that case Refs.~\cite{Pel2005JMP,Sly2006CMP})
\begin{eqnarray}                                                  
    &&\text{rot}\textbf{E}
    =-\dfrac{1}{c}\dfrac{\partial\textbf{H}}{\partial t},
    \qquad
    \text{div}\textbf{E}=4\pi\sigma,\nonumber
    \\
    &&\text{rot}\textbf{H}
    =\dfrac{1}{c}\dfrac{\partial\textbf{E}}{\partial t}
    -\dfrac{4\pi}{c}\textbf{j},~
    \text{div}\textbf{H}=0,\label{eq:3.6}
\end{eqnarray}
where the quantities~$\sigma(\textbf{x},t)$ and
$\textbf{j}(\textbf{x},t)$ are the mean values of the charge and
current densities (see~(\ref{eq:2.3})--(\ref{eq:2.5})),
\begin{equation}\label{eq:3.7}                                    
    \textbf{j}(\textbf{x},t)
    =\text{Sp}\hat{\rho}(t)\hat{\textbf{j}}(\textbf{x}),\quad
    \sigma(\textbf{x},t)
    =\text{Sp}\hat{\rho}(t){\hat\sigma}(\textbf{x}),
\end{equation}
and the quantities~$\textbf{E}(\textbf{x},t)$ and
$\textbf{H}(\textbf{x},t)$ are the electric and magnetic field
intensities, which can be expressed in terms of the potentials of an
external electromagnetic field
(see~Refs.~\cite{Akhiezer1981,Sly2006CMP}). Note that the mean
values of the quantities~$\sigma(\textbf{x},t)$ and
$\textbf{j}(\textbf{x},t)$ should be calculated in accordance with
Eq.~(\ref{eq:3.1}).  For the equilibrium state, characterized by the
statistical operator~(\ref{eq:3.1}), the mean values of these
quantities vanish (see below).

Omitting rather cumbersome calculations, we come to the following
formulas for the mean values of the charge and current densities
that are induced by the weak external electromagnetic field (see
Ref.~\cite{Sly2006CMP} for more details):
\begin{eqnarray}                                                  
    &&\sigma(\textbf{x},t)=
    \int\limits_{-\infty}^{\infty}dt'\int d^{3}x'\biggl(
    -G_{i}(\textbf{x}-\textbf{x}',t-t')\nonumber
    \\
    &&\quad\times\dfrac{1}{c}A_{i}(\textbf{x}',t')
    +G(\textbf{x}-\textbf{x}',t-t')\varphi(\textbf{x}',t')
    \biggr),\label{eq:3.8}
    \\
    &&j_{k}(\textbf{x},t)=
    \int\limits_{-\infty}^{\infty}dt'\int d^{3}x'\biggl(
    -G_{kl}(\textbf{x}-\textbf{x}',t-t')\nonumber
    \\
    &&\quad
    \times\dfrac{1}{c}A_{l}(\textbf{x}',t')
    +G_{k}(\textbf{x}-\textbf{x}',t-t')
    \varphi(\textbf{x}',t')\biggr).\label{eq:3.9}
\end{eqnarray}
The Green functions, in terms of which the
expressions~(\ref{eq:3.8}) and (\ref{eq:3.9}) are written, are
defined by
\begin{eqnarray}                                                  
    &&G(\textbf{x},t)=-i\theta(t)\text{Sp}
    {w[\hat{\sigma}(\textbf{x},t),\hat{\sigma}(0)]},\nonumber
    \\
    &&G_{k}(\textbf{x},t)=-i\theta(t)\text{Sp}
    {w[\hat{j}_{k}(\textbf{x},t),\hat{\sigma}(0)]},\label{eq:3.10}
    \\
    &&G_{kl}(\textbf{x},t)=-i\theta(t)\text{Sp}
    {w[\hat{j}_{k}(\textbf{x},t),\hat{j}_{l}(0)]},\nonumber
\end{eqnarray}
where the charge and current density operators (see
Eqs.~(\ref{eq:2.4}) and (\ref{eq:2.5})) are taken in the Heisenberg
representation.

Let us recall that we consider the response of an ideal gas of
hydrogenlike atoms to a perturbation by an external electromagnetic
field. As mentioned above, this consideration is reasonable in the
case of a description of a BEC in dilute vapors of alkali-metal
atoms. Taking into account this fact, the expressions for the
Fourier transforms of the Green functions (see Eq.~(\ref{eq:3.5}))
can be written as
\begin{subequations}\label{eq:3.11}                               
\begin{eqnarray}
    G(\textbf{k},\omega)
    &=&\dfrac{1}{\mathcal{V}}
    \sum\limits_{\textbf{p}}
    \sum\limits_{\alpha,\beta}
    \sigma_{\alpha\beta}(\textbf{k})
    \sigma_{\beta\alpha}(-\textbf{k})\nonumber
    \\
    & &\times
    \dfrac{f_{\alpha}(\textbf{p}-\textbf{k})
    -f_{\beta}(\textbf{p})}
    {\varepsilon_{\alpha}(\textbf{p})-
    \varepsilon_{\beta}(\textbf{p}-\textbf{k})
    +\omega+i0},\label{eq:3.11a}
\end{eqnarray}
\begin{eqnarray}
    G_{l}(\textbf{k},\omega)
    &=&\dfrac{1}{\mathcal{V}}
    \sum\limits_{\textbf{p}}
    \sum\limits_{\alpha,\beta}
    \left(\dfrac{(2\textbf{p}-\textbf{k})}{2M}
    \sigma_{\alpha\beta}(\textbf{k})
    +\textbf{I}_{\alpha\beta}(\textbf{k})
    \right)_{l}\nonumber
    \\
    & &\times
    \dfrac{\sigma_{\beta\alpha}(-\textbf{k})
    \left[f_{\alpha}(\textbf{p}-\textbf{k})
    -f_{\beta}(\textbf{p})\right]}
    {\varepsilon_{\alpha}(\textbf{p})-
    \varepsilon_{\beta}(\textbf{p}-\textbf{k})
    +\omega+i0},\label{eq:3.11b}
\end{eqnarray}
\begin{eqnarray}
    G_{lj}(\textbf{k},\omega)
    &=&\dfrac{1}{\mathcal{V}}
    \sum\limits_{\textbf{p}}
    \sum\limits_{\alpha,\beta}
    \left(
    \dfrac{(2\textbf{p}-\textbf{k})}{2M}
    \sigma_{\alpha\beta}(\textbf{k})
    +\textbf{I}_{\alpha\beta}(\textbf{k})
    \right)_{l}\nonumber
    \\
    & &\times
    \left(
    \dfrac{(2\textbf{p}-\textbf{k})}{2M}
    \sigma_{\beta\alpha}(-\textbf{k})
    +\textbf{I}_{\beta\alpha}(-\textbf{k})
    \right)_{j}\nonumber
    \\
    & &\quad\times
    \dfrac{f_{\alpha}(\textbf{p}-\textbf{k})
    -f_{\beta}(\textbf{p})}
    {\varepsilon_{\alpha}(\textbf{p})-
    \varepsilon_{\beta}(\textbf{p}-\textbf{k})
    +\omega+i0}.\label{eq:3.11c}
\end{eqnarray}
\end{subequations}
Here $\varepsilon_{\alpha}(\textbf{p})
=\varepsilon_{\alpha}+\textbf{p}^{2}/{2m}$ is the total energy of an
atom, and $f_{\alpha}(\textbf{p})$ is the distribution function of
the ideal gas of hydrogenlike atoms in equilibrium, which is defined
by the relation
\begin{equation*}
    \text{Sp}w\hat{\eta}^{\dag}_{\alpha}(\textbf{p})
    \hat{\eta}_{\alpha'}(\textbf{p}')
    =f_{\alpha}(\textbf{p})\delta_{\alpha\alpha'}
    \Delta(\textbf{p}-\textbf{p}'),
\end{equation*}
where $\delta_{\alpha\alpha'}$ and $\Delta(\textbf{p}-\textbf{p}')$
denote the Kronecker symbols. Bearing in mind that the unperturbed
state of a gas is characterized by the Hamiltonian~(\ref{eq:2.2}),
and using Eq.~(\ref{eq:3.2}), it is easy to come to the explicit
form for the distribution function~$f_{\alpha}(\textbf{p})$,
\begin{equation}\label{eq:3.12}                                     
    f_{\alpha}(\textbf{p})=
    (\exp\{[\varepsilon_{\alpha}(\textbf{p})
    -\mu_{\alpha}]/T\}-1)^{-1}.
\end{equation}
Employing the last expression and
Eqs.~(\ref{eq:2.4})--(\ref{eq:2.7}), one may ensure that the mean
values of $\hat{\sigma}(\textbf{x})$ and
$\hat{\textbf{j}}(\textbf{x})$ with the statistical operator~$w$
(see Eqs.~(\ref{eq:3.1}) and (\ref{eq:3.2})) vanish. We recall that
such mean values are not accounted in the derivation of
Eqs.~(\ref{eq:3.8}) and (\ref{eq:3.9}). Let us also note that the
total set of the Green functions (see Eqs.~(\ref{eq:3.11})) is given
for a completeness of the description. Below we use only the
expression for the scalar Green function~(\ref{eq:3.11a}).

As to Eqs.~(\ref{eq:3.11}), it is necessary to note the following.
Strictly speaking, these formulas are correct only in the case when
the system is considered in a state of total statistical
equilibrium. Thus, to produce excited atoms with the set of quantum
numbers~$\alpha$ in this state, one needs to keep matter in
equilibrium with radiation. For that, in turn, one needs to preserve
the number of photons in the volume in which the gas of hydrogenlike
atoms is contained (or to return the photons by the special set of
\textquotedblleft mirrors'').

In the opposite case, the statistical operator~(\ref{eq:3.2})
describes a certain quasistationary state of the system, but not the
state of total statistical equilibrium. By the term
\textquotedblleft quasistationary equilibrium'' we mean a state of
the system in which atoms can be found only in quantum states with
lifetimes much longer than other characteristic times of system
relaxation (e.g., the relaxation due to collision processes). For
example, as these states we can consider the hyperfine ground states
of alkali-metal atoms (see Ref.~\cite{Sly2007LTP}) and also states
whose existence is stimulated by the external field (e.g., resonant
laser radiation). Therefore, in the case of absence of equilibrium
between radiation and matter, the sum over all quantum
states~$\alpha,\beta$ in Eqs.~(\ref{eq:3.11}) must be replaced by
the sum over these characteristic quantum states. Exactly these
states are considered below in this paper.

In terms of the derived Green function~(\ref{eq:3.11a}), one can
express the permittivity of the studied system (see in this case
also Refs.~\cite{Akhiezer1981,Sly2006CMP}) as
\begin{equation*}                                                 
    \epsilon^{-1}(\textbf{k},\omega)
    =1+\dfrac{4\pi}{k^2}G(\textbf{k},\omega).
\end{equation*}
This relation can be written in the form
\begin{equation}\label{eq:3.12-2}                                 
    \epsilon(\textbf{k},\omega)
    =1-\dfrac{4\pi k^{-2}G(\textbf{k},\omega)}
    {1+4\pi k^{-2}G(\textbf{k},\omega)}.
\end{equation}
Taking into account the explicit form~(\ref{eq:3.11a}), one finds
\begin{eqnarray}                                                  
    \epsilon^{-1}(\textbf{k},\omega)
    =1+\dfrac{4\pi}{k^2}
    \dfrac{1}{\mathcal{V}}
    \sum\limits_{\textbf{p}}
    \sum\limits_{\alpha,\beta}
    \sigma_{\alpha\beta}(\textbf{k})
    \sigma_{\beta\alpha}(-\textbf{k})\nonumber
    \\
    \times
    \dfrac{f_{\alpha}(\textbf{p}-\textbf{k})
    -f_{\beta}(\textbf{p})}
    {\varepsilon_{\alpha}(\textbf{p})-
    \varepsilon_{\beta}(\textbf{p}-\textbf{k})
    +\omega+i0}.\label{eq:3.13}
\end{eqnarray}
We must note that, in accordance with Ref.~\cite{Akhiezer1981}, the
formulas~(\ref{eq:3.11}), (\ref{eq:3.12-2}), and (\ref{eq:3.13}) can
be used only in the region of relatively large~$\textbf{k}$.
Regarding the present paper, this criterion of applicability can be
written as $\nu {\lambda}^{3}\ll 1$ ($\lambda$ is the laser
wavelength and $\nu$ is the density of atoms). In the case $\nu
{\lambda}^{3}\gtrsim1$, the mentioned expressions become incorrect.
This results from the fact that in this region the interaction
between particles may play a significant role. Note that we neglect
it in deriving Eqs.~(\ref{eq:3.11}), (\ref{eq:3.12-2}), and
(\ref{eq:3.13}). To take into account the interaction effects, one
needs to derive the equations for the Green functions and to find a
hierarchy for the correlation functions (see, e.g.,
Refs.~\cite{Akhiezer1981,Abrikosov1975,Ruo1997PRA1}). In fact, the
use of these procedures results in a redefinition of the quantity
$G(\textbf{k},\omega)k^{-2}$ in Eq.~(\ref{eq:3.12-2}) because of the
account of correlation corrections \cite{Dal1995PRA,Ruo1997PRA1}.
But, in the case when this quantity is small, Eqs.~(\ref{eq:3.11}),
(\ref{eq:3.12-2}), and (\ref{eq:3.13}) and the expressions that
result from them remain correct in the first order of perturbation
theory even if one takes the correlations into account. Hence, we
consider that the mentioned quantity in Eq.~(\ref{eq:3.12-2}) is
small in the next calculations,
\begin{equation}\label{eq:3.13.3}                                 
    |4\pi k^{-2}G(\textbf{k},\omega)|\ll 1.
\end{equation}
This inequality makes it possible to ignore the correlation
corrections to the mentioned term in the first order of perturbation
theory. We return to the discussion of this approximation below.

\begin{figure}
\includegraphics{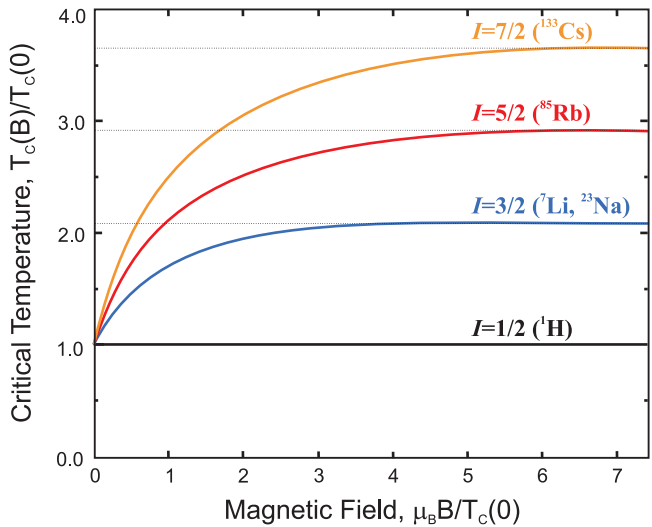}
    \caption{\label{fig:01crit}
    Dependence of critical temperature on the
    magnetic field intensity for the gas of alkali-metal atoms
    with different nuclear spins. The effect is caused by
    the Zeeman splitting of the lower hyperfine levels of
    the ground state. The gas of atoms is
    considered in the statistical equilibrium state.}
\end{figure}

As is known, at extremely low temperatures a Bose-Einstein
condensate of alkali-metal atoms can be formed. The main peculiarity
of an ideal Bose gas is that at temperatures lower than the critical
temperature, $T\leq T_\text{c}$, the chemical
potential~$\mu_{\alpha}$ does not depend on $T$ and equals the
energy of the lowest level of the atomic states that form the BEC
(see Refs.~\cite{Sly2007LTP,Akh1998JETP}),
\begin{equation*}                                           
    \mu_{\alpha}=\varepsilon_{\alpha_0},
\end{equation*}
where by the index~$\alpha_0$ we denote the set of quantum numbers
of the ground state. Let us note that this fact, in particular,
results in rather strong dependence of the critical
temperature~$T_\text{c}$ on the intensity of an external static
magnetic field  (see Fig.~\ref{fig:01crit}, and also
Ref.~\cite{Sly2007LTP}).

At temperatures much lower than $T_\text{c}$, $T\ll T_\text{c}$, we
should consider all atoms in a BEC state. Because of this fact, the
distribution function~$f_\alpha (\textbf{p})$ (see
Eq.~(\ref{eq:3.12})) is set proportional to the Dirac delta
function~$\delta(\textbf{p})$ (see, e.g., Ref.~\cite{Akhiezer1981}):
\begin{equation}\label{eq:3.13.2}                                 
    f_\alpha (\textbf{p})=(2\pi)^3\nu_\alpha\delta(\textbf{p}),
\end{equation}
where $\nu_\alpha$ is the density of condensed atoms in the
$\alpha$~state. Taking into account this fact, after integration of
(\ref{eq:3.13}) over $\textbf{p}$, the formula for the permittivity
of a gas in the BEC state ($T\rightarrow0$) takes the form
\begin{eqnarray}                                                  
    \epsilon^{-1}(\textbf{k},\omega)
    \approx 1+\dfrac{4\pi}{k^2}
    \sum\limits_{\alpha,\beta}
    \sigma_{\alpha\beta}(\textbf{k})
    \sigma_{\beta\alpha}(-\textbf{k})\nonumber
    \\
    \times\left(
    \dfrac{\nu_{\alpha}}
    {\omega+\Delta\varepsilon_{\alpha\beta}
    -\varepsilon_{\text{r}}+i\gamma_{\alpha\beta}}\right.\nonumber
    \\
    \qquad\left.-\dfrac{\nu_{\beta}}
    {\omega+\Delta\varepsilon_{\alpha\beta}
    +\varepsilon_{\text{r}}+i\gamma_{\alpha\beta}}\right),\label{eq:3.14}
\end{eqnarray}
where $\varepsilon_{\text{r}}=k^2/2m$ is the recoil energy of an
atom, and the quantity~$\sigma_{\alpha\beta}(\textbf{k})$ is still
defined by Eq.~(\ref{eq:2.4}). Here we also introduce the natural
linewidth~$\gamma_{\alpha\beta}$, which is related to the
probability of a spontaneous transition between the $\alpha$ and
$\beta$ states. Note that, in our calculations, we use the system of
units, in which the Planck constant~$\hbar$ equals unity. Thus, the
frequency is measured in energy units.

As one can conclude from the relation~(\ref{eq:3.14}), in the case
of the BEC regime, the response of the gas increases proportionally
to the density of condensed atoms. This effect takes place due to
the coherent behavior of the atoms in the condensate. Thus, one may
say that the atoms respond to the perturbation cooperatively. As for
the mathematical form, this fact corresponds to the substitution of
the wave functions of a single atom for the condensate wave
functions (see Eqs.~(\ref{eq:2.6}), and
(\ref{eq:3.13})--(\ref{eq:3.14})). In the limit of zero
temperatures, the mentioned phenomenon results in increase of
collective effects, while the value of the corresponding linewidth
(which is defined by the imaginary summand~$i\gamma_{\alpha\beta}$)
remains the same as for the isolated atom.

%

It is easy to see that at frequencies close to the energy
intervals~$\Delta\varepsilon_{\alpha\beta}$
($\Delta\varepsilon_{\alpha\beta}\equiv\varepsilon_{\alpha}
-\varepsilon_{\beta}$), some peculiarities appear in
Eq.~(\ref{eq:3.14}). Because of the small values of the recoil
energy and the natural linewidth, these peculiarities are similar to
the resonance ones. The mentioned behavior of the permittivity must
have a strong impact on the dispersion characteristics of a gas and,
as known, becomes the governing condition for slowing of the
electromagnetic pulses.

To study the propagation of electromagnetic waves in a BEC we must
supplement Eq.~(\ref{eq:3.14}) with the dispersion relation for free
waves, which can spread in the system
\begin{equation}\label{eq:3.15}                                    
    \dfrac{\omega^{2}}{c^{2}}
    \epsilon(\textbf{k},\omega)
    \mu(\textbf{k},\omega)-k^{2}=0,
\end{equation}
where $\mu(\textbf{k},\omega)$ denotes the magnetic permeability of
the gas. This quantity may also be found in the framework of the
Green-function formalism (see also Ref.~\cite{Sly2006CMP}).

\section{\label{sec.4}Dispersion characteristics of a two-level system in the BEC state}
In this section we try to simplify the description of the
interaction of electromagnetic waves with a BEC. To this end, let us
consider a case when the frequency of the external field is close to
the transition frequency between two given quantum states of atoms.
In this case (see, e.g., Ref.~\cite{Allen1987}), we can consider
only resonant terms in Eq.~(\ref{eq:3.14}), i.e., we neglect the
presence of other states of the energy spectrum of atoms and treat
the atom as a two-level system.

Let us denote by indices 1 and 2 the quantities related to the lower
and the upper states of the two-level system and omit the terms that
are related to other states. Then, Eq.~(\ref{eq:3.14}) takes the
form
\begin{equation}\label{eq:4.1}                                     
    \epsilon^{-1}(\textbf{k},\omega)
    \approx1+\dfrac{a(\textbf{k})}
    {\delta\omega+i\gamma}
\end{equation}
with
\begin{equation}\label{eq:4.4}                                  
    a(\textbf{k})=4\pi(\nu_{1}-\nu_{2})\dfrac{g_1 g_2
    |\sigma_{12}(\textbf{k})|^2}{k^2}.
\end{equation}
Here $g_j$ is the degeneracy order of the $j$ level ($j=1,2$) with
respect to total momentum,
$\delta\omega=(\omega-\Delta\varepsilon_{21})$ is the detuning of
the external field (${|\delta\omega|\ll\Delta\varepsilon_{21}}$),
and $\gamma\equiv\gamma_{12}$ is the linewidth related to the
transition probability from the upper to lower state. We also
neglect the so-called recoil energy~$\varepsilon_{\text r}$ (see
Eq.~(\ref{eq:3.14})) because of the fact that in the cases studied
below it is rather small in comparison with the characteristic
quantities that give the leading contribution to the slowing
phenomenon.

The expression~(\ref{eq:4.4}) can be written also in the more usual
form (cf. Refs.~\cite{Dal1995PRA,Ruo1997PRA1})
\begin{equation}\label{eq:4.1-2}                                  
    \epsilon(\textbf{k},\omega)
    \approx1-\dfrac{a(\textbf{k})}
    {\delta\omega+i\gamma}
    \dfrac{1}{1+C},
\end{equation}
where $C={a(\textbf{k})}/({\delta\omega+i\gamma})$. Note that the
derived expression, accurate to within a numerical factor $1/3$
before $C$, coincides with the Lorentz-Lorenz relation (see, e.g.,
Refs.~\cite{Kru2002PRA,Ruo1997PRA}). Let us emphasize that we derive
it in the framework of the microscopic approach. The
formula~(\ref{eq:4.1-2}) results directly from
Eqs.~(\ref{eq:3.13})--(\ref{eq:3.14}). Thus, it contains the
microscopic parameters of the studied system, which are calculated
in the microscopic theory.
In accordance with Eq.~(\ref{eq:3.13.2}), the
expression~(\ref{eq:4.1-2}) carries information about the response
of the ideal gas in the pure Bose-condensed state. Note that in a
rather dense condensate there also can appear additional collective
effects related to the interaction between atoms.
Depending on the value of parameter $(\nu {\lambda}^{3})$, one
should consider such effects (see
Refs.~\cite{Dal1995PRA,Ruo1997PRA1,Ruo1997PRA,Jav1999PRA}). As it is
mentioned above, the account of interaction effects in the region
$\nu {\lambda}^{3}\gg1$ results in redefining the shift~$C$. But let
us recall that we consider the case of small $C$ (see
Eq.~(\ref{eq:3.13.3})),
\begin{equation}\label{eq:4.1-3}                                  
    |{a(\textbf{k})}/({\delta\omega+i\gamma})|\ll 1.
\end{equation}
In some sense, this inequality results from the fact that we
describe the system of alkali-metal atoms as an ideal gas. This
consideration means that we neglect the interaction between atoms
(see Eq.~(\ref{eq:2.2.2})). Approximate estimates of the
applicability of this approach are given in the Appendix.

To analyze the propagation properties of electromagnetic waves in a
BEC, let us introduce the refractive index and damping factor. We
also set the magnetic permeability equal to unity in the dispersion
relation~(\ref{eq:3.15}), $\mu(\textbf{k},\omega)\approx 1$. It is
easy to verify that this approximation is valid in this case. The
refractive index~$n'(\textbf{k},\omega)$ and the damping
factor~$n''(\textbf{k},\omega)$ are expressed in terms of the
real~$\epsilon'$ and imaginary~$\epsilon''$ parts of the
permittivity~$\epsilon$, $\epsilon=\epsilon'+i\epsilon''$, as
follows:
\begin{equation}\label{eq:4.2}                                  
    n'=\dfrac{\sqrt{\sqrt{\epsilon'^2+\epsilon''^2}
    +\epsilon'}}
    {\sqrt{2}},\quad
    n''=\dfrac{\sqrt{\sqrt{\epsilon'^2+\epsilon''^2}
    -\epsilon'}}
    {\sqrt{2}}.
\end{equation}
The quantities~$\epsilon'$ and $\epsilon''$, in turn, are found from
Eq.~(\ref{eq:4.1}):
\begin{equation}\label{eq:4.3}                                  
    \epsilon'=\dfrac{\delta\omega
    (\delta\omega+a)+\gamma^2}
    {(\delta\omega+a)^2+\gamma^2},\quad
    \epsilon''=\dfrac{\gamma a}
    {(\delta\omega+a)^2+\gamma^2}.
\end{equation}

It is known that the group velocity of the electromagnetic pulse is
defined by the relation
\begin{equation}\label{eq:4.5}                                  
    v_{\text{g}}=\dfrac{c}
    {n'+\omega(\partial n'/\partial \omega)},
\end{equation}
where $c$ denotes the speed of light in vacuum. Hence, in the case
of small energy losses and strong dispersion, one finds from
Eq.~(\ref{eq:4.3})
\begin{equation}\label{eq:4.6}                                  
    v_{\text{g}}\approx2c\dfrac
    {\left((\delta\omega+a)^2+\gamma^2
    \right)^2}
    {a\omega\left[(\delta\omega+a)^2-\gamma^2
    \right]}.
\end{equation}
This expression gives the possibility of studying the dependence of
the group velocity not only on the frequency detuning~$\delta\omega$
and the linewidth~$\gamma$, but also on the characteristic
parameters of the system, which give a contribution to the value of
the parameter~$a$ (see Eq.~(\ref{eq:4.4})).

Now we define the conditions when significant slowing of
electromagnetic pulses in a two-level system with a BEC is possible
without essential absorption. Considering the
limit~${\delta\omega\rightarrow0}$ and employing Eq.~(\ref{eq:4.3}),
one gets the limits for the real and imaginary parts of the
permittivity:
\begin{equation*}                                               
    \lim\limits_{\delta\omega\rightarrow0}
    \epsilon'=\dfrac{\gamma^2}
    {\gamma^2+a^2},\quad
    \lim\limits_{\delta\omega\rightarrow0}
    \epsilon''=\dfrac{\gamma a}
    {\gamma^2+a^2}.
\end{equation*}
According to these relations, the condition of small
dissipation~($|\epsilon''|\ll\epsilon'$) may be written as
\begin{equation*}                                               
    \dfrac{|a|}{\gamma}\ll1.
\end{equation*}
Note that this inequality defines not only the limits of the
transparency region for electromagnetic waves, it also defines the
upper limits of applicability of our approach (see
Eqs.~(\ref{eq:3.13.3}) and (\ref{eq:4.1-3})). Really, as it is easy
to see, in the case ${|a|}\sim{\gamma}$, the second term in the
denominator of Eq.~(\ref{eq:4.1-2}) (the so-called Lorentz-Lorenz
shift $C$) can become of the order of unity. But, as mentioned
above, we assume that this term is small ($|C|\ll1$).

We need also to use the condition defining sufficient pulse slowing
(the strong dispersion condition), ${v_\text{g}}\ll{c}$, which, in
accordance with~(\ref{eq:4.6}), take the form
\begin{equation}\label{eq:4.7}                                  
    \dfrac{c}{v_\text{g}}\approx
    \dfrac{\Delta\varepsilon_{21}|a|}{2\gamma^2}\gg1.
\end{equation}
In a system with defined energy structure of atoms
($\Delta\varepsilon,\gamma=\text{const}$) the only parameter that
may be varied is the level occupation
difference~$(\nu_{1}-\nu_{2})$, which is included in the
parameter~$a$ (see Eq.~(\ref{eq:4.4})). Taking into account this
fact and the above inequalities, one gets
\begin{equation}\label{eq:4.8}                                  
    \dfrac{\gamma}{\Delta\varepsilon_{21}}
    \ll\dfrac{|a|}{\gamma}\ll1.
\end{equation}
This inequality characterizes the region of densities
(parameter~$a$) in which the phenomenon of significant slowing of
electromagnetic waves in a BEC may be observed.

Next, on the basis of the derived formulas, we consider two cases of
slowing of weak electromagnetic waves in a BEC. The first one is
concerned with electromagnetic pulses in the optical region of the
spectrum (resonance D~lines of alkali-metal atoms), and the second
one is related to microwaves (hyperfine transitions of the ground
states of alkali metals).

\section{\label{sec:5}Slowing of electromagnetic waves in the optical
                        range in a BEC}
As was mentioned above, a significant slowing of electromagnetic
pulses was observed in a condensed gas of sodium atoms in
Ref.~\cite{Hau1999N}. In this experiment, for the observation of
ultraslow light pulses the effect of electromagnetically induced
transparency (EIT; see Ref.~\cite{Har1997PT} for details) was used.
For realization of this effect one needs to use an additional laser
of large intensity besides the laser pulse, which is slowed in the
experiment. But we should note that the description of this
experiment in the framework of the Green-function formalism is
impossible. The main reason is that for EIT support it is necessary
to use a high-power (coupling) laser. Thus, the influence of such a
laser on a BEC cannot be described correctly within the linear
response theory. But let us emphasize that the use of a high-power
laser is not a necessary condition for the existence of the slowing
phenomenon.

Therefore, we consider the above microscopic theory in the case of
slowing of weak light pulses in a sodium BEC at temperatures~$T\ll
T_c$ in the absence of EIT. In particular, in the framework of the
developed approach we study the slowing phenomenon, when the laser
frequency is close to the transition from the ground state ($l=0$,
$3S_{1/2}$~state, $F=2$ for sodium) to the excited one ($l=1$,
$3P_{3/2}$~state, $F'=1$). Note, that the dipole
transition~$1\rightarrow2$ between these states is allowed.

First of all, we emphasize the following. The matrix element of the
charge density~$\sigma_{12}$ (see Eq.~(\ref{eq:2.4})) contains the
wave functions of a hydrogenlike atom~$\varphi_{1,2}(\textbf{y})$,
which have a sharp maximum at the point~$\textbf{y}=0$. Due to this
fact we can expand the exponents $\exp{\left[ i(m_{\text{p}}/m)
\textbf{k}\textbf{y}\right]}$ and $\exp{\left[-i(m_{\text{e}}/m)
\textbf{k}\textbf{y}\right]}$ into series of perturbation theory
over $(\textbf{ky})\ll1$. Then, in the first order of perturbation
theory, one gets
\begin{equation*}                                                           
    \sigma^{(1)}_{12}(\textbf{k})\approx i\textbf{k}\textbf{d}_{12},
\end{equation*}
where $\textbf{d}_{12}$ is the atomic dipole moment, which is
related to the dipole transition $1\rightarrow2$:
\begin{equation}\label{eq:5.1}                                             
    \textbf{d}_{12}=e \textbf{r}_{12},\quad
    \textbf{r}_{12}=\int
    d\textbf{y}~\varphi_{1}^{*}(\textbf{y})
    \varphi_{2}(\textbf{y})\textbf{y}.
\end{equation}

For the quantitative estimates we take for the sodium atom
$|\sigma_{12}(\textbf{k})|^2\approx k^2 d^2/3$ that results in (see
Eq.~(\ref{eq:4.4})) $a=(4\pi/3)(\nu_{1}-\nu_{2})d^2$. Taking
$d^2\approx(3.52er_0)^2 S_{FF'}$~\cite{Steck2000}, where $r_0$ is
the Bohr radius and $S_{FF'}$ is the relative strength of the
$F\rightarrow F'$ transition, $\Delta\varepsilon_{21}\approx
2.1$~eV, and the linewidth~$\gamma=S_{FF'}\Gamma_{e}/2$
($\Gamma_{e}=4.1\times10^{-8}$~eV is the natural
linewidth~\cite{Steck2000}; below we consider $F=2\rightarrow F'=2$
transition, $S_{22}=1/4$). Since we consider the propagation of weak
electromagnetic waves through the condensate, we assume that the
density of atoms in the excited state is negligibly small. In other
words, we treat the occupation of the excited level is small due to
low intensity of the propagating pulse ($\nu_2\ll\nu_1$). Next,
using Eq.~(\ref{eq:4.8}), one can obtain the following inequality,
which defines the conditions of slowing in condensed sodium vapor at
$T\ll T_c$:
\begin{equation}\label{eq:5.2}                                              
    3\times10^{5}\ll\nu\ll
    10^{14}~\text{cm}^{-3}.
\end{equation}
From this inequality it is easy to see that the density value of
condensed atoms in the case of real experiments with BECs
($\nu=5\times10^{12}~\text{cm}^{-3}$; see Ref.~\cite{Hau1999N}) lies
in the limits of the derived interval.

Note that the main parameter that can be measured in experimental
conditions is the relative intensity of the transmitted light. This
quantity for a homogeneous medium is defined by the expression
$I/I_{0}=\exp{[-\epsilon''(\omega)kL]}$, where $L$ is the
transversal size of the atomic sample. Taking the density of
condensed atoms in the ground state $\nu=5\times10^{11}$~cm$^{-3}$,
$k=1.066$~cm$^{-1}$, $L=0.004$~cm, one finds $I/I_{0}\approx
e^{-2.2}\approx0.11$. This allows to state that in the limit of zero
temperatures, $T\ll T_c$, ultraslow light pulses of weak intensity
may be directly observed for these parameters (according to
Eq.~(\ref{eq:4.7}), $v_{\text g}/c\approx10^{-6}$). However, a
significant part (up to 89\%) of the pulse is absorbed by the
medium.

It is easy to see that the intensity of the transmitted light has a
rather strong dependence on the system parameters, which make a
contribution to the exponent. Thus, the situation with the slowing
phenomenon may completely change, when temperature effects are taken
into account. These effects may increase the value of
$\epsilon''(\omega)$ resulting in the strong reduction of the
relative intensity of the transmitted light. For example, with
increase of $\epsilon''(\omega)$ by several times the intensity
reduces to less than 1\%. It is possible to take into account the
influence of temperature effects on the propagation properties in
the framework of the Green-function formalism. However, this is a
separate problem, which is not considered in the present paper.

Note also that in the case described above the group velocity of the
pulse is negative (anomalous dispersion in the case of a normal
population, $\nu_1>\nu_2$). We return to the discussion of this fact
in Sec.~\ref{sec:6}.

Next, let us consider light slowing in a multicomponent BEC
employing the developed theory. Nowadays, the realization of a
multicomponent Bose condensate in experimental conditions is rather
achievable (see, e.g., Ref.~\cite{Cor1998PRL}). The formulation of
this problem is interesting for the following reasons. As is known,
in an external magnetic field the Zeeman splitting of levels occurs,
and the spacing between different components of the multiplet
$\Delta\varepsilon_{\text{mag}}$ is proportional to the magnetic
field intensity. Thus, in a weak external magnetic field this energy
difference is much smaller than the difference between the nonsplit
levels of the hyperfine transition~$\Delta\varepsilon_{\text{hf}}$.
If the atoms can occupy different magnetic states of the multiplet,
we can talk about the existence of a multicomponent BEC. In such a
condensate the effects related to the resonant absorption for some
defined frequencies may be significantly smaller than in the case
described above.

\begin{figure}
\includegraphics{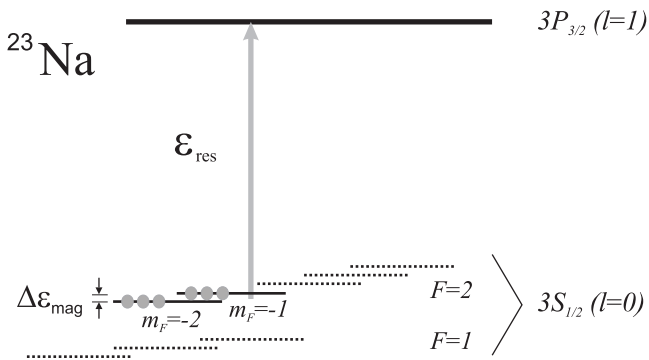}
    \caption{Scheme of energy levels
    of a sodium atom. For the sake of simplicity the levels are drawn not to
    scale, and the hyperfine structure of the excited state is not
    shown.} \label{fig:02sod}
\end{figure}

As an example, we consider a two-component condensate of sodium
atoms in the quantum states $3S_{1/2}$, $F=2$, $m_F=-2$ and $3S_{1/2}$,
$F=2$, $m_F=-1$ (below we denote them with subscripts 1 and 2,
respectively). Now, let us study the response of the system to
laser radiation, which is detuned relative to the transition between
these two states and the state~$3P_{3/2}$ (denoted below by the
subscript~3), as shown in Fig.~\ref{fig:02sod}.

\begin{figure}
\includegraphics{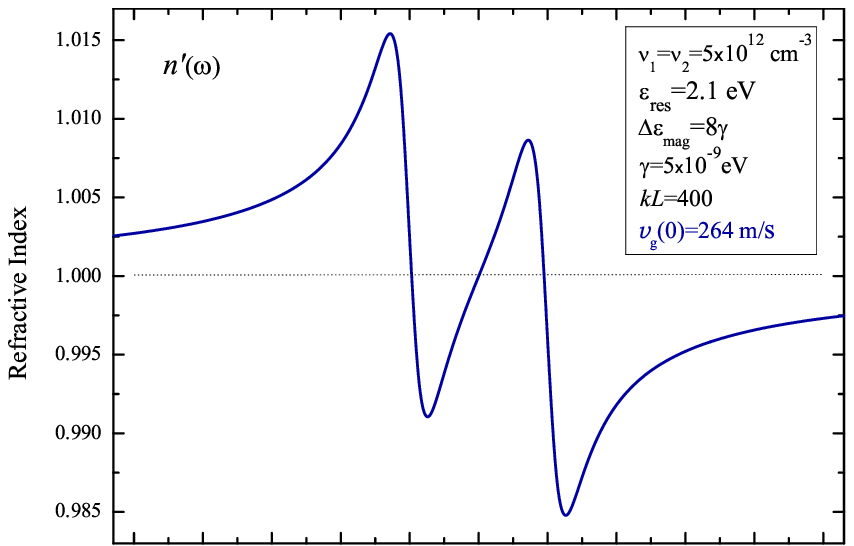}
\includegraphics{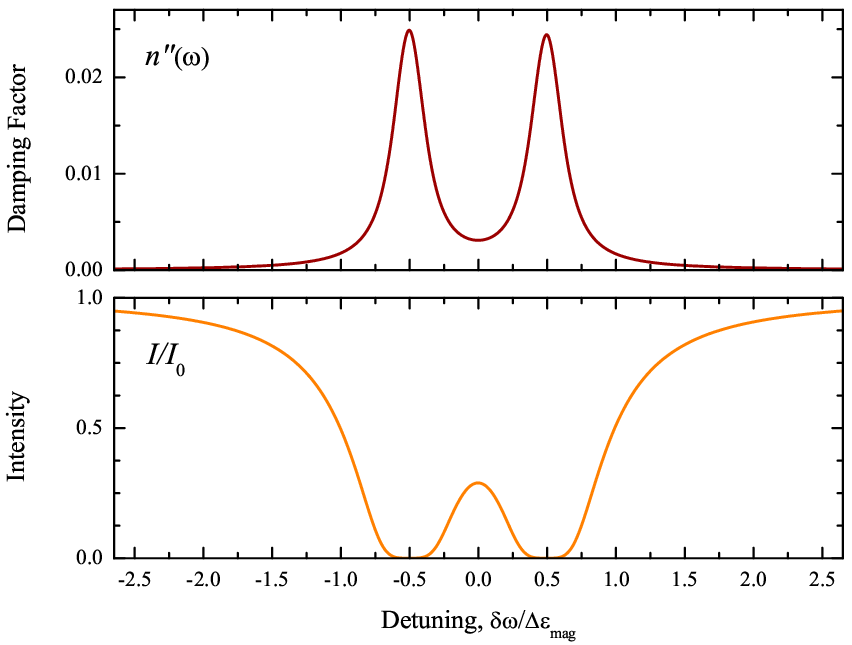}
    \caption{Dependencies of the
    refractive index, damping factor, and transmitted light intensity
    for the two-component Bose condensate.}
    \label{fig:03disp}
\end{figure}

By using Eq.~(\ref{eq:3.14}), we study the dependence of the
dispersion characteristics of an ideal gas of sodium atoms in a BEC
state on the frequency of external radiation. For simplicity, let us
treat as equal the transition probabilities from the upper state to
one of the lower states, $\gamma_{31}=\gamma_{32}$. We also consider
the densities of atoms in two lower states
$\nu_{1}=\nu_{2}=5\times10^{12}$~cm$^{-3}$ and
$\sigma_{13}=\sigma_{23}$. In fact, these assumptions are made only
for convenience, but they are not necessary in principle for
numerical evaluations. Let the energy of splitting be
$\Delta\varepsilon_{\text{mag}}=8\gamma_{31}$. Taking all other
parameters the same as in the above estimates, one finds the
dependencies of the dispersion characteristics, as illustrated in
Fig.~\ref{fig:03disp}.
From these graphs we conclude that in such a system the signal may
slow down approximately to 264~m/s.

At first sight, the last result contradicts a well-known
statement from classical optics (see, e.g., Ref.~\cite{Scully2002})
that in a two-level gas a large index of refraction is usually
related to a high absorption ratio. Hence, on atomic
resonances the signal is mostly absorbed. However, this
contradiction is only apparent. Let us emphasize one more time that in the
case described above the real part of the susceptibility
$\chi'=(\epsilon'-1)$, which gives the main contribution in the change
of the refractive index, is rather small in magnitude (as one can
conclude from Fig.~\ref{fig:03disp}, $|\chi'|\lesssim10^{-2}$). The
imaginary part $\chi''\equiv\epsilon''$, which defines the
absorption ratio, is rather small too, $\chi''\lesssim10^{-2}$ (in
the central part $\chi''\approx10^{-3}$). In other words, in the
considered case, the matter is not dense for laser radiation (the
refractive index is close to unity), but has a strong dispersion due
to closely situated resonances. This is the reason why the group
velocity of the pulse may be sufficiently reduced, while its
absorption ratio remains rather low.

Let us note that  in the regions of anomalous dispersion in
Fig.~\ref{fig:03disp} (where the derivative of the refractive index
is negative) there is good agreement with the results obtained in
Ref.~\cite{Kru2002PRA}, in which the authors studied the dependence
of the refractive index on the laser intensity. The agreement takes
place in the limit of low light intensity. In fact, we consider only
this limiting case in the present paper.

As for the obtained dependencies as whole, they are similar to the
dependencies that are observed in the EIT regime \cite{Har1997PT}.
One can, by analogy with Ref.~\cite{Hau1999N}, state that the
phenomenon of simultaneous manifestation of low group velocity value
and low absorption ratio in the central part of
Fig.~\ref{fig:03disp} is the result of destructive interference
between two absorption-emission paths, which correspond to the
transitions $1\rightarrow3$ and $2\rightarrow3$, respectively. The
difference is that the spacing between these transitions equals not
the Rabi frequency value $\Omega_{\text{R}}$, but the energy of
splitting of two occupied magnetic sublevels
$\Delta\varepsilon_{\text{mag}}$. If we compare the dependence of
light absorption (the lowest graph in Fig.~\ref{fig:03disp}) with
the analogous one in the EIT scheme \cite{Har1997PT}, we can see
that the intensities of the transmitted light in the central part
are nearly the same ($I/I_0\approx0.3$). However, in the EIT case,
the maximum is sharper due to the nonlinear nature of the effect.

Let us emphasize that in our case the energy spacing between
magnetic states must be larger than the linewidth of the
transitions. In other words, these sublevels cannot be situated
arbitrarily near each other. Therefore, the intensity of the
transmitted light depends directly on the energy spacing
$\Delta\varepsilon_{\text{mag}}$ and on the density of the condensed
states in the condensate $\nu_{1,2}$. At the same time, the group velocity of
the pulse also strongly depends on both these characteristics.
Thus, if one tries to change the characteristics of the system for much
stronger reduction of the group velocity value (increase in the density
of atoms or decrease in the energy spacing
$\Delta\varepsilon_{\text{mag}}$), the ratio of absorption will
correspondingly increase (see also Ref.~\cite{Scully2002}). In our
case, the absorption ratio is the main limiting factor for the
achievement of lower values of the group velocity.

As for the effect of electromagnetically induced transparency, in
the experiment the Rabi frequency (which is analogous to the energy
$\Delta\varepsilon_{\text{mag}}$ in our case) is usually set to be
of the same order as the linewidth. Thus, the effect of slowing due
to the Zeeman splitting is somewhat weaker ($v_{\text{g}}\approx
264$~m/s in comparison with $v_{\text{g}}\approx 32$~m/s in
Ref.~\cite{Hau1999N}). However, the authors hope that the
observation of this kind of slowing is rather possible in the
present experimental conditions.

\section{\label{sec:6}Slowing of microwaves by hyperfine ground states
            of alkali-metal atoms}

As mentioned above, the proposed microscopic theory is correct when
the occupation of the excited atomic states is stimulated by an
external electromagnetic field (the case that is studied in
Sec.~\ref{sec:5}), or when the lifetimes of the excited states
are much larger than the relaxation times. Note that for alkali-metal
atoms all sublevels of the hyperfine ground state can be treated as
such long-living states. This follows from the fact that the dipole
transitions are forbidden for these states. Waves whose frequency
corresponds to the energy difference between the levels of
hyperfine structure are usually called microwaves. Since there
are many trapping-related experiments with ultracold atoms prepared
in different hyperfine states, the problem of the possibility of slowing
microwaves becomes rather significant.

By considering the example of cesium atoms in the BEC state, we try to
find the conditions in which, in our opinion, the slowing
phenomenon may be observed experimentally. To this end, according to
the developed approach, let us show that the region
defined by Eq.~(\ref{eq:4.8}) exists. Note that a brief description
of the application of the microscopic approach for the description of
microwave slowing may be found in Ref.~\cite{Sly2008JLTP}.

It is known that alkali-metals ($^{133}$Cs, in particular) do not have a
dipole moment in the ground state. This agrees with
Eq.~(\ref{eq:5.1}), from which one can find that the dipole
moment~$\textbf{d}$ equals zero for the wave functions of the ground
state. Thus, to describe the slowing of microwaves on the levels of
hyperfine structure with forbidden dipole transitions, we should
expand the charge density matrix element~$\sigma_{12}$ (see
Eq.~(\ref{eq:2.4})) to the second order in~$(\textbf{ky})\ll1$. As a
result, one gets
\begin{equation}\label{eq:6.1}                                     
    \sigma^{(2)}_{12}(k)\approx\dfrac{e}{3}(kr_0)^2,
\end{equation}
where $r_0$ is the atomic radius (for the cesium ground state
$r_0\approx 2.6\times10^{-8}$~cm \cite{Cle1963JCP}). Let us consider
that the levels are degenerate, $g_1=7$, $g_2=9$ ($g_{j}=2F_{j}+1$, where
$F_j$ is the total spin of an atom in the $j$ hyperfine state, $j=1,2$),
and also set $k=(\Delta\varepsilon_{21}/c)$ with
$\Delta\varepsilon_{21}\approx 3.8\times10^{-5}$~eV (microwaves with
frequency 9.2~GHz). We take the natural linewidth~$\gamma$ of the
upper hyperfine level equal to the value that correspond to the
anticipated accuracy ($10^{-16}$) in \textquotedblleft cesium fountain clock''
experiments~\cite{Sal1999PRL}, $\gamma\approx3.8\times10^{-21}$~eV.
Then, on the basis of Eqs.~(\ref{eq:4.4}) and (\ref{eq:4.8}), one may
get the following region where the slowing phenomenon may be observed:
\begin{equation}\label{eq:6.2}                                     
    10^{-1}\ll|\nu_1 - \nu_2|\ll
    10^{15}~\text{cm}^{-3}.
\end{equation}
It is easy to see that the effect becomes greater when the
density difference increases. When it reaches the upper limit of
(\ref{eq:6.2}), damping effects prevail in the system and the
signal is mostly absorbed by the medium. We stress that the region of
densities~(\ref{eq:6.2}), in our opinion, looks convenient from the
standpoint of experiments with the BEC regime in cesium (see, e.g.,
Ref.~\cite{Web2003S}, in which cesium atoms with the
density~$\nu=1.3\times10^{13}~\text{cm}^{-3}$ were used).

Now we note the following. As it is easy to see from
Eq.~(\ref{eq:4.6}), in the limit~$\delta\omega\rightarrow0$, the sign
of the group velocity~$v_{\text g}$ depends directly on the sign of
the quantity~$a$, which, in turn, depends on the sign of the
difference~$(\nu_1 - \nu_2)$. In other words, it depends on whether
the population is normal or inverse. In the case of normal population
$(\nu_1 > \nu_2)$, the group velocity of the signal is negative. It is
traditionally considered that the group velocity for the transparent
matter is positive, and it can be negative only in regions in which
the pulse is mostly absorbed. But, from our theory one can conclude
that, according to Eq.~(\ref{eq:4.8}), the signal can propagate in
the system with rather small dissipation (i.e., the matter is
transparent) and rather slow velocity. Note that the observation of
electromagnetic pulses with negative group velocity is not so
abnormal. The existence of this kind of phenomenon for physical
systems when the wave frequency is close to atomic (or molecular)
resonances was pointed out and studied in many works (both
theoretical~\cite{Gar1970PRA,Cri1971PRA} and
experimental~\cite{Chu1982PRL,Mac1985PL}). In the case of inverse
population $(\nu_1 < \nu_2)$, more \textquotedblleft usual'' situation takes place
because the group velocity of the slowed pulse is positive.

\begin{figure}
    \includegraphics{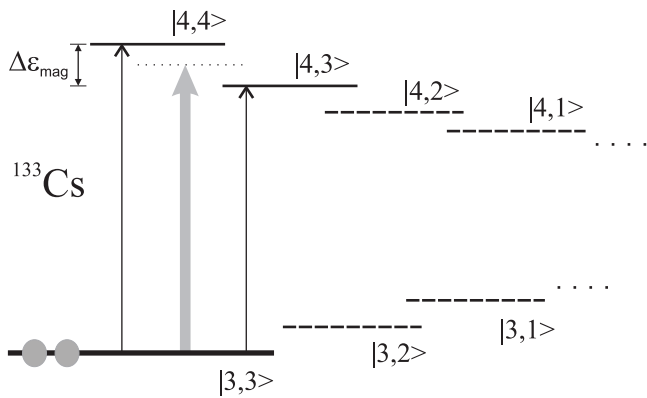}
    \caption{Level scheme of cesium atom in an external magnetic
    field. The first sign in brackets corresponds to the total
    spin~$F$, the second one to its projection~$m_{F}$. For the sake of
    visibility, the figure is not drawn to scale and only the most
    relevant levels are shown.}
    \label{fig:04ces}
\end{figure}

So far, we assumed that the external magnetic field is absent, i.e.,
we treated the hyperfine levels as degenerate in the quantum
number~$m_F$. However, an external magnetic field (such situation
occurs in most experiments with BECs) results in a more complicated
picture. Each hyperfine level splits to additional sublevels with
different total spin projections~$m_{F}$ (e.g., for the cesium
atoms, see Fig.~\ref{fig:04ces}). As is known, in a weak (and also
in a strong) external magnetic field the energy spacing between such
sublevels is proportional to the magnetic field intensity (the
Zeeman and Paschen-Back effects, respectively).

Let us assume that the magnetic sublevels of the upper level of
hyperfine structure are kept away from each other (the linewidth of
the transition is much smaller than the energy difference between
sublevels~$\Delta\varepsilon_{\text{mag}}$, as shown in
Fig.~\ref{fig:04ces}). Then, a signal tuned exactly to the
transition between the occupied lower state and one of the upper
states (narrow black arrowed lines in Fig.~\ref{fig:04ces}) may be
slowed down, as described above. The only difference from the
previous situation is that the states are no longer degenerate
($g_{1}=g_{2}=1$) and the linewidth~$\gamma_{j}$ in
Eq.~(\ref{eq:4.1}) corresponds to the transition probability from
the chosen sublevel with the projection~$j=(-F',...,F')$ to the
occupied lower state. Taking all other values the same when deriving
the inequality~(\ref{eq:6.2}), one can find the inequality in the
presence of the external magnetic field
\begin{equation}\label{eq:6.3}                                     
    10\ll|\nu_1 - \nu_2|\ll
    10^{17}\text{cm}^{-3}.
\end{equation}
In the experiment~\cite{Web2003S}, condensed cesium atoms with a
peak density of $1.3\times10^{13}~\text{cm}^{-3}$ were kept in a
trap. Thus, according to Eq.~(\ref{eq:6.3}), one finds that in such
a system ultraslow microwaves may be observed. Moreover, direct
calculations show that the group velocity of the signal (see
Eq.~(\ref{eq:4.6})) propagating in the mentioned system is close to
$3\times10^{-4}$~cm/s.

Note that here we must make important remarks relating to the
possibility of experimental observation of ultraslow microwaves in a
BEC. In fact, it is impossible to make a direct measurement of the
group velocity value in the conditions of a real experiment. It is
usually calculated by measuring the time of pulse delay.
The characteristic dimensions of the pulse (and, specifically, the
wavelength) must be of the order of the atomic sample size, in order for it
to propagate in the system. In the present experimental conditions
this relation is not always satisfied. Usually, the characteristic
sizes of a condensate are of the order of a millimeter (or even
submillimeter), whereas the wavelength, related to the hyperfine
transition, is of the order of centimeters (e.g., for cesium in the
absence of an external field~$\lambda_\text{hf}=3.3$~cm).

In view of the aforesaid, there are two possible ways to avoid this
problem. The first one, which is rather difficult at this moment,
is to use other configurations of magneto-optical traps (and,
probably, other cooling techniques) that may allow production of
condensed clouds with centimeter sizes. The second one, and
probably more realistic, is to use a rather strong bias field, which
results (according to the Paschen-Back effect) in increase of the
spacing between levels and, consequently, in the reduction of the
resonance wavelength. At the present moment, such fields are
completely accessible, although they must be rather strong. For
example, to reduce the clock transition wavelength to
$\lambda_\text{hf}=2$~mm in cesium, the field intensity must be of
the order of 45~kG, which is quite achievable from the standpoint of
the trapping-related experiments (see Ref.~\cite{Gue2005PRL}). Let
us note also that, corresponding to the results~\cite{Sly2007LTP}
(see also Fig.~\ref{fig:01crit}), the presence of the additional
magnetic field may sufficiently raise the value of the critical
temperature.

Note that, in order to observe the effect for the limiting case
considered while deriving Eq.~(\ref{eq:4.8}), one needs to use a
microwave signal with detuning~$\delta\omega_{\text{s}}$ much
smaller than the level linewidth:
\begin{equation}\label{eq:6.4}                                     
    \delta\omega_{\text{s}}\ll\gamma_{j}.
\end{equation}
In a real experiment it is rather difficult to to satisfy such a
condition. This is due to the fact that the natural
linewidth~$\gamma_{j}$ of the hyperfine structure levels is rather
small (dipole transitions are forbidden and the transitions come
from the higher order effects).

In this case, the detuning of a maser between close-lying magnetic
sublevels of the hyperfine ground state structure may save the
situation. At the same time, we must satisfy the condition
\begin{equation*}                                                  
    \gamma_{j}\ll\delta\omega_{\text{s}}\ll
    \Delta\varepsilon_{\text{mag}},
\end{equation*}
where $\Delta\varepsilon_{\text{mag}}$ is the energy difference
between these sublevels. For example, a microwave signal that is
detuned relative to the transitions between two neighboring states of
the upper hyperfine multiplet and the occupied lower state (wide
gray arrowed line in Fig.~\ref{fig:04ces}) may satisfy this condition
at certain intensities of the magnetic fields.

\begin{figure}
\includegraphics{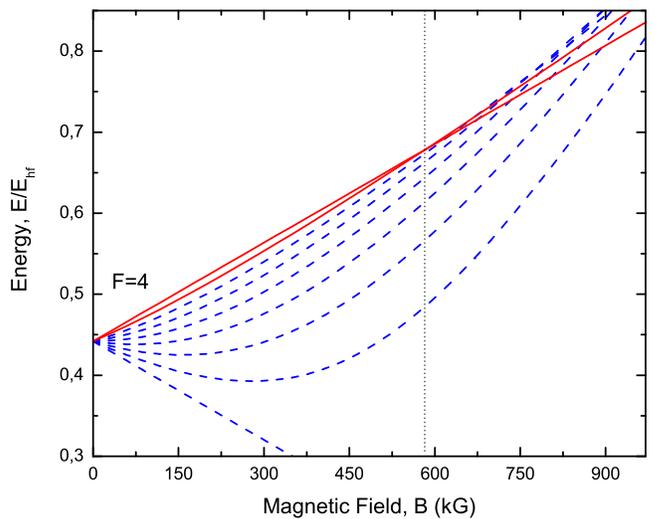}
    \caption{Energy spectrum of the upper ($F$=4)
    hyperfine multiplet of the $^{133}$Cs ground state.
    The region of intersection of the two upper
    levels corresponds to the
    intensity~$B_{\text{cr}}\approx583$~kG.} \label{fig:05lev}
\end{figure}

Let all atoms be condensed in the lower (occupied) state with the
density~$\nu$ and energy~$\varepsilon_0$. Then, it is not
difficult to come to the following expression for the permittivity
(see Eqs.~(\ref{eq:3.14}) and (\ref{eq:4.4})):
\begin{equation}\label{eq:6.5}                                     
    \epsilon^{-1}(\textbf{k},\omega)
    \approx 1+
        a\sum\limits_{j=-F'}^{F'}\left(
    \dfrac{1}
    {\omega-\Delta\varepsilon_{j}(B)
    +i\gamma_{j}}\right),
\end{equation}
where $\Delta\varepsilon_{j}=(\varepsilon_{|F',m_{F'}\rangle}
-\varepsilon_0)$. The derived expression defines the dependence of
the refractive index on the frequency of the microwave radiation.
Therefore, it also defines the conditions for pulse slowing (see
Eqs.~(\ref{eq:4.2}) and (\ref{eq:4.5})). It is easy to see that in this
case the condition of wave slowing~(cf. Eq.~(\ref{eq:4.8}) takes the
form
\begin{equation}\label{eq:6.6}                                     
    \dfrac{\Delta\varepsilon_{\text{mag}}}
    {\Delta\varepsilon_{\text{hf}}}
    \ll \dfrac{4a}{\Delta\varepsilon_{\text{mag}}}
    \ll \dfrac{\Delta\varepsilon_{\text{mag}}}{2\gamma}.
\end{equation}
The structure of the $^{133}$Cs upper hyperfine state in the presence of
magnetic field is schematically shown (not to scale) in
Fig.~\ref{fig:05lev}. From this graph, one may conclude that the
energy difference~$\Delta \varepsilon_{\text{mag}}$ is small either
in the region of weak magnetic fields or in the region of levels
intersection (strong magnetic fields).

In the presence of an external magnetic field the dependence of the
refractive index has a rather complicated form due to the large number
of magnetic states (nine sublevels of the upper $F=4$ state; see
Fig.~\ref{fig:05lev}). Hence, it is more convenient to show such
dependence in a figure.

\begin{figure}
\includegraphics{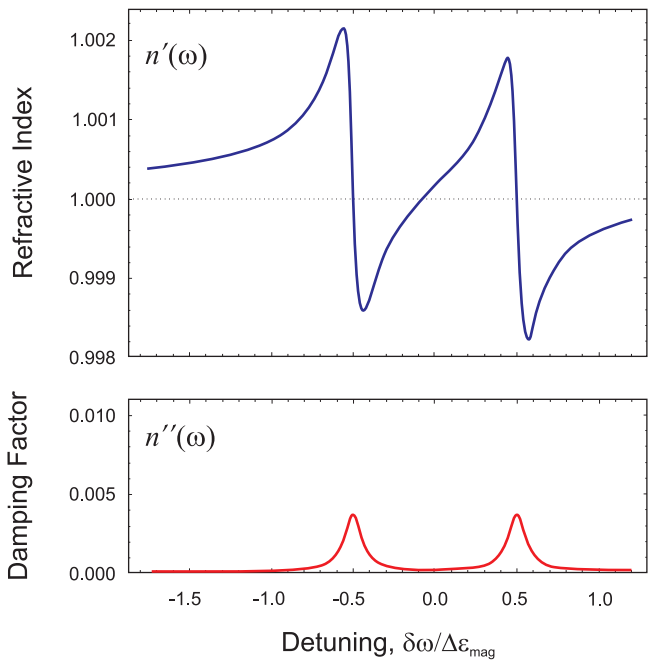}%
    \caption{Refractive index and damping factor dependencies for
    cesium atoms in the BEC state in the region of frequencies close to the
    transitions between magnetic sublevels of hyperfine splitting. Left
    and right peaks correspond to the transitions
    $|3,3\rangle\rightarrow|4,-4\rangle$ and
    |$|3,3\rangle\rightarrow|4,-3\rangle$, respectively. The
    parameter used for the Zeeman splitting energy
    $\Delta\varepsilon_{\text{mag}}/ \gamma_{j}=15$.}
\label{fig:06disp2}
\end{figure}

The dependencies of the refractive index and damping factor on the
frequency are shown in Fig.~\ref{fig:06disp2}. Notice that in the
central part ($\delta\omega=0$, corresponding to the wide gray
arrowed line in Fig.~\ref{fig:04ces}) the refractive index has a
slope with steepness dependent on the
interval~$\Delta\varepsilon_{\text{mag}}
=\varepsilon_{|4,4\rangle}-\varepsilon_{|4,3\rangle}$, which, in turn, depends
on the magnetic field intensity. This means that a microwave signal that
is detuned in this way may be sufficiently slowed. Moreover,
its group velocity depends directly on the magnetic field intensity.

The direct calculations show that, e.g., for an ideal gas of
condensed cesium atoms ($\nu=1.3\times10^{13}~\text{cm}^{-3}$,
$\gamma_{j}=3.8\times10^{-21}$~eV) with the energy of splitting
$\Delta\varepsilon_{\text{mag}}=5.7\times10^{-20}$~eV
($\Delta\varepsilon_{\text{mag}}/\gamma_{j}=15$, such as shown in
Fig.~\ref{fig:06disp2}), the pulse can be slowed down to 0.01~cm/s.
But we emphasize that in this case (weak bias field) the pulses with
wavelengths of the order of centimeters may be slowed in a BEC.
Thus, one must use condensed clouds of the same size, which is, as
mentioned above, hardly achievable from the standpoint of present
experimental conditions.

In the case of strong magnetic fields (close to the region of
intersection of levels of the Paschen-Back splitting; see
Fig.~\ref{fig:05lev}), the graphs of the refractive index and
damping factor have qualitatively the same dependence, as shown
in Fig.~\ref{fig:06disp2}. Moreover, the wavelength of the
corresponding radiation in these conditions is of the order of
submillimeters, which is quite acceptable from the standpoint of a real
experiment. But, as follows from the estimates that are based on
Eqs.~(\ref{eq:6.5}) and (\ref{eq:6.6}), the conditions for slowing of
electromagnetic waves are satisfied only in a region of magnetic
field intensity~$B_{\text{cr}}$ that approximately equals
583~kG. In this region of bias field one can effectively control the
ultraslow microwave phenomenon in a BEC. Present scientific papers
note the achievement of ultrastrong intensities of magnetic
fields (e.g., of the order of 600~kG; see in that case
Ref.~\cite{Wos2007JMMM}), but the authors of the present work, being
theorists, cannot state whether it is possible to use such fields
in BEC-related experiments.

\section{Conclusion}

Thus, we propose a microscopic approach for the description of the
linear response of an ideal gas of hydrogenlike atoms in the BEC state
to perturbation by an external electromagnetic field. The use
of this approach allows us to find the permittivity of dilute gases of
alkali-metal atoms in a BEC state as a function of the microscopic
characteristics of the system at temperatures~$T\ll T_\text{c}$.

The main advantage of the considered approach is that it comes from
a microscopic description, which is based on the approximate
formulation of the second quantization method for systems that
consist of bound states of particles \cite{Pel2005JMP}. Another
advantage is its convenience for studying the
dispersion characteristics of gases regardless of the type and
number of quantum states in the system. As a possible drawback
one can refer the fact that in the form in which it is formulated
in this work it cannot be correctly used for the description of the
response to a strong perturbation (as may probably be done in
the approaches \cite{Kru2002PRA,Jav1999PRA}) owing to the use of
linear theory. In other words, this approach cannot be used without
considerable modification to systems in which strong laser
radiation is used (e.g., the coupling field for the provision of
EIT). Note that we neglected also the interaction between atoms (in
view of the consideration of rather dilute gases) and the influence of
temperature effects. Temperature effects in the theory developed may
be taken into account, but this represents a separate problem.

The developed approach is demonstrated on the example of the
reduction of the speed of light in a BEC of sodium vapor with parameters
close to the experimental ones~\cite{Hau1999N}. It is shown that in
some conditions the signal may be sufficiently slowed in the absence
of an additional strong (coupling) field. A strong field is used only
to provide electromagnetically induced transparency leading to
the reduction of the damping factor value. Thus, it is not a
necessary condition for the observation of ultraslow pulses.

The following cases are considered: the optical signal is tuned
to the resonance frequency and the signal is detuned relative to
close transitions in the two-component condensate. It is shown that
in these situations ultraslow pulses may be observed. The
conditions for slowing, which depend on the macroscopic
characteristics of the system, are found. The dispersion
dependencies are studied also.

The approach is applied for studying the propagation of microwaves
in a BEC. It is shown that the microwave signal, which is tuned to
the transition between different hyperfine levels of the ground
state, can propagate in the system with rather small energy loss. It
is pointed out that the sign of the group velocity of the propagating
pulse depends directly on the occupation difference of the states.

The influence of an external static and homogeneous magnetic field
(taking into account the Zeeman and Paschen-Back effects) on the
slowing parameters is studied. Considering cesium atoms, it is shown
that the pulse can be slowed down to extremely small values in some
conditions.


\appendix

\section{\label{app.A}On the correctness of the ideal gas approximation}

Let us now say some words relating to the limits of applicability of
the developed approach. Note that we consider the
condensed gas of alkali-metal atoms as ideal. In other words, we assume that
the interaction between atoms is small in comparison with the
binding energy and in comparison with the interaction of particles
with the external field.

To estimate the influence of effects related to the interatomic
interaction, let us consider the Hamiltonian
$\hat{\mathcal H}_{\text{int}}$, which is introduced in
Eq.~(\ref{eq:2.1}). As mentioned above, the microscopic
expressions for it are found in Ref.~\cite{Pel2005JMP} in the
framework of the developed formulation of
the second-quantization method. It is also shown that
in the low-energy approximation (or in the case of low temperatures)
this operator can be written as
\begin{equation}\label{eq:2.2.2}                                    
    \hat{\mathcal H}_{\text{int}}=\int d\textbf{x}~d\textbf{y}~
    v_{\alpha\beta;\gamma\delta}(\textbf{x}-\textbf{y})
    {\hat\eta}_{\alpha}^{\dag}(\textbf{x})
    {\hat\eta}_{\beta}^{\dag}(\textbf{y})
    {\hat\eta}_{\gamma}(\textbf{y})
    {\hat\eta}_{\delta}(\textbf{x}),
\end{equation}
where the Einstein summation convention over indices $\alpha,\beta,
\gamma,\delta$ is meant, and the kernel
$v_{\alpha\beta;\gamma\delta}(\textbf{z})$ has the form
\begin{equation}\label{eq:2.2.2b}                                    
    v_{\alpha\beta;\gamma\delta}(\textbf{z})
    =\dfrac{1}{z^5}\left[z^2(\textbf{d}_{\alpha\delta}\textbf{d}_{\beta\gamma})
    -3(\textbf{z}\textbf{d}_{\alpha\delta})(\textbf{z}\textbf{d}_{\beta\gamma})
    \right],
\end{equation}
in which the quantities~$\textbf{d}_{\alpha\beta}$ are the matrix
elements of the dipole moment operator of a single atom. Such matrix
elements can be expressed in terms of the atomic wave functions
$\varphi_{\alpha}(\textbf{y})$ in different quantum states $\alpha,
\beta$,
\begin{equation}\label{eq:2.2.3}                                     
    \textbf{d}_{\alpha\beta}=e\int
    d\textbf{y}~\varphi_{\alpha}^{*}(\textbf{y})
    \varphi_{\beta}(\textbf{y})\textbf{y}.
\end{equation}
Thus, one can see that in this case the Hamiltonian~(\ref{eq:2.2.2})
corresponds to the dipole-dipole interaction of atoms.

It is easy to verify that in the case of the interaction of atoms in the
ground state the expression~(\ref{eq:2.2.2}) is inapplicable. It
results from the fact that the dipole moment of the hydrogenlike
atom in the ground state is equal to zero. One may find the
Hamiltonian of interaction ${\hat{\mathcal H}}^{(0)}_{\text{int}}$
of the atoms in the ground state from the second-order terms of the
perturbation theory over the interaction~(\ref{eq:2.2.2}). If we assume
that this interaction is small in comparison with the binding
energy~(\ref{eq:2.2}), we get (see Ref.~\cite{Pel2005JMP})
\begin{equation}\label{eq:2.2.4}                                     
    \hat{\mathcal H}^{(0)}_{\text{int}}
    =\dfrac{1}{2}
    \int d\textbf{x}~d\textbf{y}~
    v^{(0)}(\textbf{x}-\textbf{y})
    {\hat\eta}_{\alpha_0}^{\dag}(\textbf{x})
    {\hat\eta}_{\alpha_0}^{\dag}(\textbf{y})
    {\hat\eta}_{\alpha_0}(\textbf{y})
    {\hat\eta}_{\alpha_0}(\textbf{x}),
\end{equation}
where ${\hat\eta}_{\alpha_0}^{\dag}(\textbf{x})$ and
${\hat\eta}_{\alpha_0} (\textbf{x})$ are the creation and
annihilation operators of hydrogenlike (alkali-metal) atoms in the ground
state with the set of quantum numbers~$\alpha_0$ at the
point~$\textbf{x}$, respectively. The quantity~$v^{(0)}(\textbf{z})$
can be expressed by the relation
\begin{equation}\label{eq:2.2.5}                                     
    v^{(0)}(\textbf{z})=
    \dfrac{{\sum\limits_{\beta\lambda}}^{\prime}
    |v_{\beta\lambda;\alpha_0\alpha_0}(\textbf{z})|^{2}}
    {{2\varepsilon_{\alpha_0}-
    \varepsilon_{\beta}-\varepsilon_{\lambda}}},
\end{equation}
where $v_{\beta\lambda;\alpha_0\alpha_0}(\textbf{z})$ is defined by
Eq.~(\ref{eq:2.2.2b}), and $\varepsilon_{\alpha_0}$ is the
ionization energy of an atom. The prime in the sum means that the
terms with $\beta=\alpha_0$ and $\lambda=\alpha_0$ are omitted.
Hence, one can conclude that the Hamiltonian~(\ref{eq:2.2.4})
corresponds to the well-known Hamiltonian of the van der Waals
interaction (see, e.g., Ref.~\cite{Landau1977}).

We must note that it is convenient to replace the
operators~(\ref{eq:2.2.2}) and (\ref{eq:2.2.4}) by effective
Hamiltonians (see in this case Refs.~\cite{Abrikosov1975,Pitaevskii2003}).
Thus, the effective
Hamiltonian of the dipole-dipole interaction takes the form
\begin{equation}\label{eq:2.2.6}                                     
    \hat{\mathcal H}_{\text{int}}^{\text{(eff)}}
    =U_{\alpha\beta;\gamma\delta}\int d\textbf{x}~
    {\hat\eta}_{\alpha}^{\dag}(\textbf{x})
    {\hat\eta}_{\beta}^{\dag}(\textbf{x})
    {\hat\eta}_{\gamma}(\textbf{x})
    {\hat\eta}_{\delta}(\textbf{x})
\end{equation}
with (see also Ref.~\cite{Bethe1977})
\begin{equation}\label{eq:2.2.7}                                     
U_{\alpha\beta;\gamma\delta}\equiv{4\pi\over
3}\textbf{d}_{\alpha\delta}\textbf{d}_{\beta\gamma}.
\end{equation}
In the case of interaction of atoms in the ground state, the
Hamiltonian~(\ref{eq:2.2.4}) can be replaced by
\begin{equation}\label{eq:2.2.8}                                     
    \hat{\mathcal H}^{(0)}_{\text{int}}={1\over 2}
    U_{0}\int d\textbf{x}~
    {\hat\eta}_{\alpha_0}^{\dag}(\textbf{x})
    {\hat\eta}_{\alpha_0}^{\dag}(\textbf{x})
    {\hat\eta}_{\alpha_0}(\textbf{x})
    {\hat\eta}_{\alpha_0}(\textbf{x}),
\end{equation}
where
\begin{equation}\label{eq:2.2.9}                                     
    U_{0}\equiv
    \dfrac{4\pi{\hbar}^{2}a_{\text{s}}}{m}
\end{equation}
and $a_{\text{s}}$ is the $s$-wave scattering length (see, e.g.,
Ref.~\cite{Pitaevskii2003}). Note that the quantities
$U_{\alpha\beta;\gamma\delta}$ and $U_{0}$ may be defined by Eqs.~(\ref{eq:2.2.7}) and (\ref{eq:2.2.9}) only in the case
when the quantum fluctuations, which are related to the
elementary excitations in the system, are negligibly small \cite{Pitaevskii2003}.

To find the approximate criteria of applicability of our theory, let
us consider a dilute weakly nonideal Bose gas, which interacts with
a weak external electromagnetic field. It is known that the
standard thermodynamic perturbation theory is inapplicable to such a
gas in the region below the critical point~$T_\text{c}$. It results
from the fact that there appear diverging terms in the region of
small momenta (see in this case, e.g., Refs.~\cite{Akhiezer1981,
Pitaevskii2003}). Therefore, one needs to use a special perturbation
theory. Such a theory was constructed by Bogoliubov in
\cite{Bog1963OIY}. Within this theory it is shown (and rigorously
proven in Ref.~\cite{Bogoliubov1970}) that the creation and
annihilation operators may be considered as \emph{c} numbers,
\begin{equation*}                                                          
    {\hat\eta}_{\alpha_0}^{\dag}(\textbf{x})
    \rightarrow\nu_{\alpha}e^{-\phi},\quad
    {\hat\eta}_{\alpha_0}(\textbf{x})
    \rightarrow\nu_{\alpha}e^{\phi},
\end{equation*}
where $\phi$ is some phase, which is irrelevant in our
consideration. To account for the existence of the overcondensate
particles (the density of which is much smaller than the density of
the condensate particles) one needs to introduce a gas of elementary
excitations, or quasiparticles with an energy spectrum, which is
now called the Bogoliubov spectrum. The special perturbation theory,
which was proposed by Bogoliubov, removes the divergences of the
correlation functions in the region of small
wave vectors \cite{Pitaevskii2003}. In addition to the mentioned
spectrum, this theory allows one also to find the ground state energy of
the weakly nonideal Bose gas, its chemical potential (which has a
nonzero value in case of interaction), and the corrections to these
values, which result from the existence of quantum fluctuations.
For example, the contribution of interaction effects to the ground state
energy~${\cal E}_0$ can be written as
(see Ref.~\cite{Pitaevskii2003})
\begin{equation}\label{eq:5.3}                                              
    {\cal E}_{0}=U{N^2\over2V}
    \left(1+\dfrac{128}{15\sqrt{\pi}}
    (\nu a_{\text{s}}^3)^{1/2}\right),
\end{equation}
where $U$ is the physical coupling constant (see Eqs.~(\ref{eq:2.2.7}) and
(\ref{eq:2.2.9})), $N$ is the overall number of atoms in the weakly
nonideal Bose gas, $V$ is the system volume, $\nu$ is the density of
particles, and $a_{\text{s}}$ is the scattering length (see
Eq.~(\ref{eq:2.2.9})). The second summand in brackets in
Eq.~(\ref{eq:5.3}) characterizes the quantum fluctuation
contribution to the ground state energy of the gas. The
formula~(\ref{eq:5.3}) is correct, when the so-called gas
parameter~$\nu |a_{\text{s}}|^3$ is small,
\begin{equation}\label{eq:5.4}                                              
    \nu |a_{\text{s}}|^3\ll 1.
\end{equation}
Below we assume that the relation~(\ref{eq:5.4}) is satisfied.
Therefore, in the following considerations we do not take into account the
contribution of the quantum correlations to the characteristics of
the weakly nonideal Bose gas interacting with the external
radiation.

We note that the formalism developed in the present
paper can be applicable to the system under consideration
only in the case when the interatomic interaction is small
compared to the interaction with the external field. Thus,
let us compare the energy related to the interaction with the weak
external radiation and the contribution of the interaction between
atoms to the ground state energy~${\cal E}_{0}$. If we neglect the
existence of the quantum fluctuations in the system, this
contribution, in
accordance with Eq.~(\ref{eq:5.3}), takes the form
\begin{equation}\label{eq:5.8}                                              
    {\cal E}_{0}=
    V\sum_{\alpha}U_{\alpha}
    {\nu_{\alpha}\nu_{\alpha_0}},
\end{equation}
where $U_{\alpha}$ is the coupling constant, which characterizes
the effective Hamiltonian of interaction between atoms with the set
of quantum numbers $\alpha$ (see
Eqs.~(\ref{eq:2.2.6})--(\ref{eq:2.2.9})), and $\nu_{\alpha}$ is the
density of such atoms in the condensate (by the index~$\alpha_0$ we
denote the set of quantum numbers of the ground state). The energy
related to the interaction with the external electromagnetic field
is characterized by the Hamiltonian~$\hat{V}(t)$ (see
Eq.~(\ref{eq:2.3})). Its mean value~${\cal E}_{E}$ can be expressed
in the framework of the Green-function formalism (see, e.g.,
Refs.~\cite{Akhiezer1981,Sly2006CMP}). But, one can estimate it also in a
different way. This energy must be of the same order of magnitude as
the energy, which is pumped into the system by the external field
\cite{Akhiezer1981}. As mentioned above, the
expression for the energy that is absorbed by the system from the
external field can be written as (see Eq.~(\ref{eq:4.3}))
\begin{equation}\label{eq:5.9}                                              
    {\cal E}_{E}\sim V{E^2\over
    8\pi}\xi,
\end{equation}
where $E$ is the amplitude of the electric field in the pulse,
$\xi=\{1-\exp{[-\epsilon''(\omega)kL]}\}$ is the part of the energy
that is absorbed by the system in the case of propagation of a
single pulse, and $L$ is the characteristic linear size of the
atomic cloud.

The ideal gas approximation is correct when the energy ${\cal
E}_{0}$ is small in comparison with ${\cal E}_{E}$,
\begin{equation}\label{eq:5.10}                                            
    \sum_{\alpha}U_{\alpha}{\nu_{\alpha}
    \nu_{\alpha_0}}\ll
    \dfrac{E^2}{8\pi}
    \xi.
\end{equation}
In Sec.~\ref{sec.4} we study the dispersion characteristics of a
two-level system in the BEC state. There we consider that the density of
atoms in the excited state (denoted by the set of quantum numbers
$\alpha_1$) is much less than  the density of atoms in the ground
state (denoted by the set of quantum numbers $\alpha_0$),
$\nu_{\alpha_1}\ll\nu_{\alpha_0}$. The number of atoms in the
excited states can be estimated with the assumption that all
energy~(\ref{eq:5.9}) that is absorbed by the system is spent on
the excitation of certain states $\alpha_1$. If the intensity of the
electromagnetic field is rather small, one gets (see, e.g.,
Ref.~\cite{Svelto1982})
\begin{equation}\label{eq:5.11}                                            
    \nu_{\alpha_1}\sim
    \dfrac{E^2}{\hbar\omega}
    \xi.
\end{equation}
From the inequality $\nu_{\alpha_1}\ll\nu_{\alpha_0}$ we have
\begin{equation}\label{eq:5.12}                                            
    \dfrac{E^2}{\hbar\omega}\xi
    \ll\nu_{\alpha_0}.
\end{equation}
Therefore, using Eq.~(\ref{eq:5.11}) we come to the following
relation for the contribution of the interaction (\ref{eq:5.8}) in
the ground state energy (see also Eqs.~(\ref{eq:2.2.7}),
(\ref{eq:2.2.9})):
\begin{equation*}                                                          
    {\cal E}_{0}={4\pi{\hbar}^{2}a_{\text{s}}\over m}\nu_{\alpha_0}^{2}
    +{4\pi\over3}|\textbf{d}_{\alpha_0 \alpha_1}|^{2}\nu_{\alpha_0}
    {E^2\over\hbar\omega}\xi.
\end{equation*}
Thus, the inequality (\ref{eq:5.10}) can be written as
\begin{equation}\label{eq:5.13}                                           
    {4\pi{\hbar}^{2}a_{\text{s}}\over m}\nu_{\alpha_0}^{2}+{4\pi\over
    3}|\textbf{d}_{\alpha_0 \alpha_1}|^{2}\nu_{\alpha_0}
    {E^2\over\hbar\omega}\xi
    \ll{E^2\over 8\pi}\xi.
\end{equation}
This relation defines the limits for the intensity of the external
electromagnetic field and the density
of condensate particles in the ground state.

To find numerical estimates, we take the value of the field intensity
from the experiment~\cite{Hau1999N}, in which a laser with
power density 5~mW/cm$^{2}$ and $\lambda=$589~nm was used. The
scattering length~$a_{\text{s}}$ for a number of alkali-metals
in the ground state is of the
order of a few nanometers \cite{Pitaevskii2003}. Taking the
density $\nu_{\alpha_0}\sim10^{13}~ \text{cm}^{-3}$ and
$\xi\approx0.5$, one easily obtains
\begin{eqnarray}                                                           
    {4\pi{\hbar}^{2}a_{\text{s}}\over m}\nu_{\alpha_0}^{2}
    &\sim&10^{-12}~\text{erg~cm}^{-3},\nonumber
    \\
    {4\pi\over
    3}|\textbf{d}_{\alpha_0 \alpha_1}|^{2}\nu_{\alpha_0}
    {E^2\over\hbar\omega}\xi
    &\sim&10^{-14}~\text{erg~cm}^{-3},\nonumber
    \\
    {E^2\over 8\pi}\xi
    &\sim&10^{-6}~\text{erg~cm}^{-3}.\nonumber
\end{eqnarray}
It is easy to see that in this case the inequality~(\ref{eq:5.13})
is correct. Let us note that the relation~(\ref{eq:5.12}) is
satisfied also with a rather large \textquotedblleft reserve''.

Now we can conclude that the relations (\ref{eq:3.13.3}) and
(\ref{eq:4.1-3}) (see also Eqs.~(\ref{eq:5.2}) and (\ref{eq:6.3})) and
the inequalities (\ref{eq:5.10}), (\ref{eq:5.12}), and (\ref{eq:5.13})
define the limits of applicability of our theory for the description
of the response of a condensed gas of alkali-metals to a perturbation by
a weak external field. For completeness of the description, one
should also add the condition
\begin{equation*}
    {E^2\over 8\pi}\xi
    \ll\varepsilon_{\alpha_{0}}\nu_{\alpha_0}^{2},
\end{equation*}
which characterizes the smallness of the energy related to the
interaction with the external field in comparison with the
characteristic energy that may be enough to ionize atoms in the
condensate.

Note also that here we do not consider the interaction of closely
located radiating dipoles. This can be explained by the fact that the
effects produced by this interaction are quadratic in the density of
atoms in the excited state (dipoles), and the density of dipoles in
dilute vapors of alkali-metals is low (the average distance between
dipoles is much larger than  the characteristic wavelength).
First, the experimentalists \cite{Hau1999N} come to such a
conclusion. Second, one can come to it after the following
consideration. As for excited states produced by external
pumping, the statement about low density of dipoles can be proven by
estimates based on Eq.~(\ref{eq:5.11}). Note also that some part
of the excited atoms can be produced by collisions in the system.
But for a steady-state system the density of such atoms can be
comparable only in the case when there is an equilibrium between the
radiation (photons) and matter (see Ref.~\cite{Akhiezer1981}). For a
BEC in dilute vapors of alkali-metals, in which the ultraslow light
phenomenon is observed, such equilibrium can be achieved, e.g., by a
special set of mirrors that returns photons to the system. In this
case, the interaction of closely located radiating dipoles must be
taken into account because it can have a strong impact on the phenomenon
under study. The mentioned effect should be taken into account also for
rather dense gases in the BEC state.

Of course, all of the estimates introduced in this
section are rather approximate. Probably more explicit
estimates can be made only from an approach related to the
Green-function formalism (see, e.g.,
Refs.~\cite{Abrikosov1975,Ruo1997PRA1}).

\bibliography{5-PRA}

\end{document}